\documentclass[paper,11pt]{JHEP}
\pdfoutput=1
\usepackage[centertags]{amsmath}
\usepackage{amsfonts} \usepackage{amssymb} \usepackage{amsthm}
\usepackage{graphicx}

\newcommand{\ii}{\mathrm{i}}
\newcommand{\ee}{\mathrm{e}}
\newcommand{\one}{{\rm 1\kern -.9mm l}}

\newcommand{\tr}{\mathrm{tr}\,}

\newcommand{\taup}{\theta^\alpha}
\newcommand{\ldaup}{\lambda^{\dot\alpha}}
\newcommand{\lda}{\lambda_{\dot\alpha}}
\newcommand{\wda}{\bar w_{\dot\alpha}}
\newcommand{\wdb}{\bar w_{\dot\beta}}
\newcommand{\ta}{\theta_\alpha} 
\title{Non-perturbative effective interactions from fluxes
}
\author{\parbox{11.5cm}{Marco Bill\`o$^1$, Livia Ferro$^{1,2}$, Marialuisa Frau$^1$, Francesco Fucito$^3$,
Alberto Lerda$^4$ and Jose F. Morales$^3$}
\\
~\\
~\\
$^1$Dipartimento di Fisica Teorica, Universit\`a di Torino\\
and I.N.F.N. - sezione di Torino \\
Via P. Giuria 1, I-10125 Torino, Italy\\
\vspace{0.3cm}
$^2$Laboratoire de Physique Th\'{e}orique\\
\'{E}cole Normale Sup\'{e}rieure\\
24, rue Lhomond, F-
75231 Paris Cedex 05, France\\
\vspace{0.3cm}
$^3$I.N.F.N. - sezione di Roma II
\\
Via della Ricerca Scientifica, I-00133 Roma, Italy\\
\vspace{0.3cm}
$^4$Dipartimento di Scienze e Tecnologie Avanzate, Universit\`a del Piemonte Orientale\\
and I.N.F.N. - Gruppo Collegato di Alessandria - sezione di Torino\\
Via V. Bellini 25/G, I-15100 Alessandria, Italy\\
\vspace{0.3cm}
\email{billo,ferro,frau,lerda@to.infn.it, Francesco.Fucito,Francisco.Morales@roma2.infn.it}
}

\abstract{Motivated by possible implications on the problem of moduli stabilization and
other phenomenological aspects, we study D-brane instanton effects in flux compactifications.
We focus on a local model and compute non-perturbative interactions
generated by gauge and stringy instantons
in a ${\mathcal N} = 1$ quiver theory with gauge group 
$\mathrm{U}(N_0)\times \mathrm{U}(N_1)$ and matter in the bifundamentals. This model
is engineered with fractional D3-branes at a $\mathbb{C}^3/(\mathbb{Z}_2\times \mathbb{Z}_2)$  singularity and its non-perturbative sectors are described by introducing fractional D-instantons. 
We find a rich variety of instanton-generated interactions, ranging from superpotentials
and Beasley-Witten like multi-fermion terms to non-supersymmetric flux-induced instanton interactions.
}
\preprint{DFTT/20/2008\\ROM2F/2008/20\\LPTENS 08/41\\}

\begin{document}
\section{Introduction and motivations}
\label{intro}
Recently a lot of attention has been devoted to the study of
four dimensional compactifications of
Type II string theories with systems
of intersecting or magnetized D-branes that preserve $\mathcal N=1$ supersymmetry  \cite{Blumenhagen:2005mu,Blumenhagen:2006ci,Marchesano:2007de}.
These compactifications provide, in fact, promising scenarios for phenomenological applications
and realistic model building in which gauge interactions
similar to those of the supersymmetric extensions of the Standard Model of particle physics
are engineered using space-filling D-branes that partially or totally wrap
the internal six-dimensional space.
The effective actions of such brane-world models describe interactions
of gauge degrees of freedom, associated to open strings, with gravitational fields, 
associated to closed strings, and have the
generic structure of $\mathcal N=1$ supergravity in four dimensions
coupled to vector and chiral multiplets.
As is well-known \cite{Cremmer:1982en}, four-dimensional ${\mathcal N}=1$ supergravity theories
are specified by the choice of a gauge group ${\mathcal G}$,
with the corresponding adjoint fields and gauge kinetic functions, by a K\"ahler potential $K$ 
and a superpotential $W$, which are, respectively, a real and a holomorphic function 
of some chiral superfields $\Phi^i$. 
The supergravity vacuum is parametrized by the expectation values of these chiral
multiplets that minimize the scalar potential
\begin{equation}
V\,=\,\ee^{K}\left( D_i \bar W D^i W - 3 \,|W|^2\right) +D^a D_a
\end{equation}
where $D_i W\equiv\partial_{\Phi^i} W + \big(\partial_{\Phi^i} K \big)\,W$ is the K\"ahler
covariant derivative of the superpotential and the $D^a$ ($a=1,\ldots,{\rm dim}(\mathcal G)$)
are the D-terms. Supersymmetric vacua, in particular, correspond to those
solutions of the equations 
$\partial_{\Phi^i} V=0$  satisfying the D- and F-flatness conditions $D^a=D_i W=0$.

When we consider Type IIB string theory on a Calabi-Yau three-fold in presence of D3-branes,
which is the case discussed in this paper, the chiral superfields
$\Phi^i$ comprise the fields $U^r$ and $T^m$ that
parameterize the deformations of the complex and K\"ahler structures of the three-fold,
the axion-dilaton field
\begin{equation}
\label{tau}
\tau=C_0+\ii \,\ee^{-\varphi}~,
\end{equation} 
where $C_0$ is the R-R scalar and $\varphi$ the dilaton, and also some multiplets
$\Phi_{\mathrm{open}}$ coming from the open strings attached to the D-branes.
The resulting low-energy ${\mathcal N}=1$ supergravity model has 
a highly degenerate vacuum.
One way to lift (at least partially) this degeneracy is provided by the addition
of internal 3-form fluxes of the bulk theory \cite{Grana:2005jc,Douglas:2006es,Denef:2007pq}
via the generation of a superpotential
\cite{Gukov:1999ya,Taylor:1999ii}
\begin{equation}
W_{\mathrm{flux}}=\int G_3\wedge \Omega~.
\label{wflux3}
\end{equation}
Here $\Omega$ is the holomorphic $(3,0)$-form of the Calabi-Yau three-fold and 
$G_3={F}-\tau H$ is the complex 3-form flux given in terms of the R-R and NS-NS fluxes  
${F}$ and $H$. The flux superpotential (\ref{wflux3}) depends explicitly on $\tau$ 
and implicitly on the complex structure parameters $U^r$ which specify $\Omega$.
Insisting on unbroken $\mathcal N=1$ supersymmetry requires the flux $G_3$ to be an imaginary anti-selfdual 3-form of type (2,1) \cite{Giddings:2001yu}, since 
the F-terms $D_{U^r} W_{\mathrm{flux}}$,
$D_\tau W_{\mathrm{flux}}$ and $D_{T^m} W_{\mathrm{flux}}$
are proportional to the $(1,2)$, $(3,0)$ and $(0,3)$
components of the $G$-flux respectively. These F-terms can also be
interpreted as the ``auxiliary'' $\theta^2$-components of the kinetic functions
for the gauge theory defined on the space-filling branes, and thus are soft supersymmetry 
breaking terms for the brane-world effective action. Such soft terms have been computed
in various scenarios of flux compactifications
and their effects, like for instance induced masses for the gauginos and the
gravitino, have been analyzed relying on the structure of the bulk supergravity Lagrangian and on
$\kappa$-symmetry considerations (see for instance the reviews \cite{Grana:2005jc,Douglas:2006es,Denef:2007pq} and references therein) 
and recently also by a direct world-sheet analysis in \cite{Billo:2008sp}.

Beside fluxes, also non-perturbative contributions \cite{Witten:1995gx,Douglas:1995bn}
to the effective actions may play an important r\^ole in the moduli stabilization process
\cite{Denef:2005mm,Lust:2005dy} and bear phenomenologically relevant implications for
string theory compactifications.
In the framework we are considering, non-perturbative sectors are described by
configurations of D-instantons or, more generally, by wrapped Euclidean branes
which may lead to the generation of a non-perturbative superpotential of the form 
\begin{equation}
W_{\mathrm{n.p.}}= \sum_{ \{k_A\} } c_{\{k_A\}}(\Phi^i) \,\ee^{2\pi \ii\sum_A 
\!k_A \tau_A}~.
\label{wfluxnp}
\end{equation}
Here the index $A$ labels the cycles wrapped by the instantonic branes, while
$k_A$ denotes the instanton number and $\tau_A$ the complexified gauge coupling 
of a D-brane wrapping the cycle $A$.
Finally, $c_{\{k_A\}}(\Phi^i)$ are some (holomorphic) functions
of the chiral superfields whose particular form depends on the details of the
model. In general, the $\tau_A$'s depend
on the axion-dilaton modulus $\tau$ and the K\"ahler parameters $T^m$
that describe the volumes of the cycles around which the D-branes are wrapped%
\footnote{The explicit  dependence of $\tau_A$ on $\tau$ and $T^m$ 
can be derived from the Dirac-Born-Infeld action as explained in Appendix
\ref{apDBI}.}. We remark that Eq. (\ref{wfluxnp}) holds both for gauge and stringy
instantons corresponding, respectively, 
to the cases where the cycle $A$ is occupied or not by a gauge D-brane. 

The interplay of fluxes and non-perturbative contributions, leading to a combined
superpotential 
\begin{equation}
 \label{Wtot}
W = W_{\rm flux} + W_{\rm n.p.}~,
\end{equation}
offers new possibilities for finding supersymmetric vacua. Indeed,
the derivatives $D_{U^r} W_{\mathrm{flux}}$,
$D_\tau W_{\mathrm{flux}}$ and $D_{T^m} W_{\mathrm{flux}}$   
might now be compensated by $D_{U^r} W_{\mathrm{n.p.}}$,
$D_\tau W_{\mathrm{n.p.}}$ and $D_{T^m} W_{\mathrm{n.p.}}$ \cite{Lust:2005dy}
so that also the $(1,2)$, $(3,0)$ and $(0,3)$
components of $G_3$ may become compatible with supersymmetry and help in removing the vacuum
degeneracy \cite{Lust:2006zg}.

Another option could be to arrange things in such a way to have a Minkowski vacuum with $V=0$
and broken supersymmetry. If the superpotential is divided into an observable and a hidden sector,
with the flux-induced supersymmetry breaking happening in the latter, 
this could be a viable model for supersymmetry breaking mediation.
If all moduli are present in $W$, the number of equations necessary to satisfy the
extremality condition for $V$ seems sufficient to obtain a complete moduli stabilization.
To fully explore these, or other, possibilities, it is crucial however 
to develop reliable techniques to compute non-perturbative corrections to the effective action
and determine the detailed structure of the non-perturbative superpotentials that can be generated,
also in presence of background fluxes.

In the last few years there has been much progres in the analysis of non-perturbative effects 
in brane-world models and concrete computational tools have been developed using systems 
of branes with different boundary conditions \cite{Green:2000ke,Billo:2002hm}.
These methods not only allow to
reproduce \cite{Billo:2002hm}\nocite{Billo:2006jm,Akerblom:2006hx,Billo:2007sw}-\cite{Billo:2007py}
the known instanton calculus of (supersymmetric) field theories \cite{Dorey:2002ik}, but can 
also be generalized to more exotic configurations where field theory methods are not yet available \cite{Blumenhagen:2006xt}\nocite{Ibanez:2006da,Florea:2006si,Bianchi:2007fx,Argurio:2007qk,
Argurio:2007vqa,Bianchi:2007wy,Ibanez:2007rs,Antusch:2007jd,Blumenhagen:2007zk,Aharony:2007pr,
 Blumenhagen:2007bn,Camara:2007dy,Ibanez:2007tu,GarciaEtxebarria:2007zv,Petersson:2007sc,
 Bianchi:2007rb,Blumenhagen:2008ji,Argurio:2008jm,Cvetic:2008ws,Kachru:2008wt,GarciaEtxebarria:2008pi}-
 \cite{Buican:2008qe}.
The study of these exotic instantons has led to interesting results
in relation to moduli stabilization, (partial) supersymmetry breaking and even fermion masses and 
Yukawa couplings \cite{Blumenhagen:2006xt,Ibanez:2006da,Blumenhagen:2007zk}. 
A careful analysis of the moduli of this kind of instantons is however required in order
to be sure that unwanted neutral fermionic zero-modes are either absent,
as in some orientifold models \cite{Argurio:2007qk,Argurio:2007vqa,Bianchi:2007wy},
or lifted \cite{Blumenhagen:2007bn,Petersson:2007sc}. 
If really generated, such exotic interactions
could also become part of a scheme in which the supersymmetry breaking 
is mediated by non-perturbative soft-terms arising in the hidden sector of the theory, as recently 
advocated also in \cite{Buican:2008qe}. Nonetheless, the stringent conditions required for the 
non-perturbative terms to be different from zero, severely limit the freedom to
engineer models which are phenomenologically viable. 

To make this program more realistic, in this paper we address the study of the generation of non-perturbative terms in presence of fluxes. Indeed fluxes not only lead to the perturbative
superpotential (\ref{wflux3}) but also lift some zero-modes of the instanton background and allow for new types of non-perturbative couplings. In the following we will consider the 
interactions generated by gauge and stringy instantons in a specific 
setup consisting of fractional D3-branes 
at a $\mathbb C^3/(\mathbb Z_2\times \mathbb Z_2)$ singularity which engineer 
a ${\mathcal N}=1$ $\mathrm{U}(N_0)\times \mathrm{U}(N_1)$ quiver gauge theory
with bi-fundamental matter fields.
In order to simplify the treatment, still keeping the desired supergravity interpretation,
this quiver theory can thought of as a local description of a Type IIB
Calabi-Yau compactification on the toroidal orbifold 
$T^6/(\mathbb Z_2\times \mathbb Z_2)$. From this local standpoint, it is not necessary
to consider global restrictions on the number $N_0$ and $N_1$ of D3-branes, which can therefore be 
arbitrary, nor add orientifold planes for tadpole cancelation. In such a setup we then 
introduce background fluxes of type $G_{(3,0)}$ and $G_{(0,3)}$, and study the
induced non-perturbative interactions in the presence of gauge and stringy
instantons which we realize by means of fractional D-instantons. 
In this way we are able to obtain a very rich class of non-perturbative effects
which range from "exotic" superpotentials terms to non-supersymmetric
multi-fermion couplings.
We also show that, as anticipated in \cite{Billo:2008sp}, stringy instantons in presence of $G$-fluxes can generate non-perturbative
interactions even for $\mathrm{U}(N)$ gauge theories. This has to be compared with the case without
fluxes where an orientifold projection \cite{Argurio:2007qk,Argurio:2007vqa,Bianchi:2007wy} 
is required in order to solve the problem of the neutral fermionic zero-modes.
Notice also that since the $G_{(3,0)}$ and $G_{(0,3)}$ components of the $G_3$ are related 
to the gaugino and gravitino masses (see for instance \cite{Camara:2003ku,Camara:2004jj}), 
the non-perturbative flux-induced
interactions can be regarded as the analog of the Affleck-Dine-Seiberg (ADS) superpotentials 
\cite{Affleck:1983mk} for gauge/gravity theories with soft supersymmetry breaking terms.
In particular the presence of a $G_{(0,3)}$ flux has no effect on the gauge theory 
at a perturbative level but it generates new instanton-mediated effective interactions  \cite{GarciaEtxebarria:2007zv}.

For the sake of simplicity most of our computations will be carried out
for instantons with winding number $k=1$; however we also briefly discuss some multi-instanton effects.
In particular from a simple counting of zero-modes we find that in our quiver gauge theory 
an infinite tower of D-instanton corrections can contribute to 
the low-energy superpotential, even in the field theory limit with no fluxes, in constrast to what
happens in theories with simple gauge groups where the ADS-like superpotentials are generated only by instantons with winding number $k=1$. These multi-instanton effects in the quiver theories certainly 
deserve further analysis and investigations.

The plan of the paper is the following: in Section \ref{secn:d3d-1} 
we review the D-brane setup in the orbifold $\mathbb C^3/(\mathbb Z_2 \times \mathbb Z_2)$
in which our computations are carried out. 
In Section \ref{sec:effint} we discuss a quick method to infer the structure of the non-perturbative
contributions to the effective action based on dimensional analysis and symmetry considerations.
In Section \ref{secn:1inst} we analyze the ADHM instanton action and
discuss in detail the one-instanton induced interactions
in SQCD-like models without introducing $G$-fluxes.
Finally in Sections \ref{secn:fluxeffects} and \ref{secn:strinst} we consider gauge and stringy instantons in presence of $G$-fluxes and compute the non-perturbative interactions they produce.
Some more technical details are contained in the Appendix.

\section{D3/D(--1)-branes on $\mathbb{C}^3/\big(\mathbb{Z}_2\times \mathbb{Z}_2\big)$ }
\label{secn:d3d-1}

In this section we discuss the dynamics of the D3/D$(-1)$ brane system on the orbifold $\mathbb{C}^3/\big(\mathbb{Z}_2\times \mathbb{Z}_2\big)$ where the elements of
$\mathbb Z_2 \times \mathbb Z_2$ act on the three complex coordinates $z^I$ of $\mathbb C^3$
as follows
\begin{equation}
\begin{aligned}
h_1\,&: (z^1,z^2,z^3) \to (z^1,-z^2,-z^3)
~,\\
h_2\,&: (z^1,z^2,z^3) \to (-z^1,z^2,-z^3)
~,\\
h_3\,&: (z^1,z^2,z^3) \to (-z^1,-z^2,z^3)
~.
\end{aligned}
\label{hi}
\end{equation}
This material is well-known; nevertheless, we review it mainly with the purpose of setting
our notations.

\subsection{The gauge theory}
\label{subsec:gauge}
A stack of $N$ D3-branes in flat space gives rise to a four-dimensional $\mathrm U(N)$ gauge
theory with $\mathcal N=4$ supersymmetry. Its field content, corresponding to the massless excitations
of the open strings attached to the D3-branes, can be organized into
a $\mathcal N=1$ vector multiplet $V$ and three $\mathcal N=1$ chiral multiplets $\Phi^I$ ($I=1,2,3$).
These are $N\times N$ matrices:
\begin{equation}
\big\{ V,\Phi^I\big\}^u_{~v}
\label{gauge0}
\end{equation}
with $u,v,\ldots=1,\ldots,N$. In $\mathcal N=1$ superspace notation,
the action of the $\mathcal N=4$ theory is
\begin{equation}
S= \frac{1}{4\pi}\,\mathrm{Im}\left[\tau\!\!\int\!\!d^4x\, d^2\theta\, d^2 \bar \theta ~
\mathrm{Tr}\big(\bar\Phi_I\,\ee^{2V}\Phi^I\big) + \tau
\!\!\int \!\!d^4x\, d^2\theta ~\mathrm{Tr}\Big(\frac{1}{2}W^\alpha W_\alpha +
\frac{1}{3!}\epsilon_{IJK}\Phi^I \Phi^J\Phi^K\Big) \right]
\label{actgauge0}
\end{equation}
where $\tau$ is the axion-dilaton field (\ref{tau}) and $W_\alpha=-\frac{1}{4}\bar D_{\dot\alpha}
\bar D^{\dot\alpha} D_\alpha V$ is the chiral superfield whose lowest component is the
gaugino.

When the D3-branes are placed in the $\mathbb{C}^3/\big(\mathbb{Z}_2\times \mathbb{Z}_2\big)$ 
orbifold, the supersymmetry of the gauge theory is reduced to $\mathcal N=1$ and only
the $\big(\mathbb{Z}_2 \times \mathbb{Z}_2\big)$-invariant components of $V$ and $\Phi^I$
are retained. Since $V$ is a scalar under the internal $\mathrm{SO}(6)$ group, while the chiral multiplets $\Phi^I$ form a vector, it is immediate
to find the transformation properties of these fields under the orbifold group elements $h_I$. 
These are collected in Tab. \ref{tablez2z2}, where in the first columns 
we have displayed the eigenvalues of
$h_I$ and in the last column we have indicated
the $\mathbb{Z}_2\times \mathbb{Z}_2$
irreducible representation $R_A$ ($A=0,1,2,3$) under which each field transforms%
\footnote{Quantities carrying an index $A$ of the chiral or anti-chiral spinor representation of $\mathrm{SO}(6)$, like for example the gauginos $\Lambda^{\alpha A}$ or
$\bar\Lambda_{\dot\alpha A}$ of the $\mathcal N=4$ theory,
transform in the representation $R_A$ of the orbifold group; thus there is a one-to-one
correspondence between the spinor indices of $\mathrm{SO}(6)$ and those labeling the irreducible
representations of $\mathbb Z_2 \times \mathbb Z_2$; for this reason
we can use the same letters $A,B,\ldots$ in the two cases (see for example Ref.
\cite{Billo:2008sp} for details).\label{foot:1}}.
\begin{table}[ht]
\centering
\begin{tabular}{cccccc}
\hline\hline
  fields & $h_0$ & $h_1$ & $h_2$  & $h_3$ & Rep's \\
    \hline
   V & + & + & + & + & $R_0$\\
   $\Phi^1$ & +& $+$ & $-$ & $-$ & $R_1$ \\
   $\Phi^2$ & +&$-$ & + & $-$ & $R_2$ \\
   $\Phi^3$ & +&$-$ &$-$ & + & $R_3$ \\
   \hline\hline
 \end{tabular}
 \caption{$\mathbb{Z}_2\times \mathbb{Z}_2$ eigenvalues of D3/D3 fields}
 \label{tablez2z2}
 \end{table}

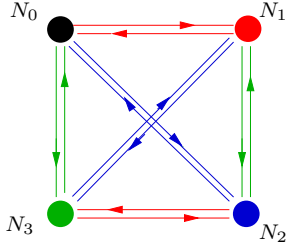
\begin{figure}[hbt]
\begin{center}
\begin{picture}(0,0)%
\includegraphics{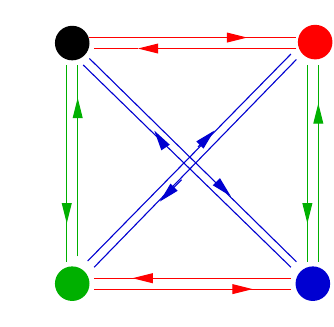}%
\end{picture}%
\setlength{\unitlength}{1381sp}%
\begingroup\makeatletter\ifx\SetFigFontNFSS\undefined%
\gdef\SetFigFontNFSS#1#2#3#4#5{%
  \reset@font\fontsize{#1}{#2pt}%
  \fontfamily{#3}\fontseries{#4}\fontshape{#5}%
  \selectfont}%
\fi\endgroup%
\begin{picture}(4559,4450)(211,-4247)
\put(4726,-136){\makebox(0,0)[lb]{\smash{{\SetFigFontNFSS{8}{9.6}{\familydefault}{\mddefault}{\updefault}$N_1$}}}}
\put(4726,-4111){\makebox(0,0)[lb]{\smash{{\SetFigFontNFSS{8}{9.6}{\familydefault}{\mddefault}{\updefault}$N_2$}}}}
\put(301,-136){\makebox(0,0)[lb]{\smash{{\SetFigFontNFSS{8}{9.6}{\familydefault}{\mddefault}{\updefault}$N_0$}}}}
\put(226,-3961){\makebox(0,0)[lb]{\smash{{\SetFigFontNFSS{8}{9.6}{\familydefault}{\mddefault}{\updefault}$N_3$}}}}
\end{picture}%
\end{center}
\caption{The quiver diagram encoding the field content and the charges for
fractional D-branes of the orbifold
$\mathbb{C}^3/(\mathbb{Z}_2\times\mathbb{Z}_2)$. The dots represent the branes associated with the irrep $R_A$
of the orbifold group. A stack of $N_A$ such branes supports a $\mathrm{U}(N_A)$ gauge theory.
An oriented link from the $A$-th to the $B$-th dot
corresponds to a chiral multiplet transforming in the
$\big({N_A},{\overline N_B}\big)$ representation of the gauge group and in the $R_A \otimes R_B$ representation of the orbifold group.}
\label{fig:quiver}
\end{figure}

To each representation $R_A$ of $\mathbb Z_2 \times \mathbb Z_2$
one associates a fractional D3 brane type. Let $N_A$ be the number
of D3-branes of type $A$ with
\begin{equation}
 \sum_{A=0}^3 N_A = N
\label{N}
\end{equation}
so that the $N\times N$ adjoint fields $V$, $\Phi^I$ of the parent theory break into
$N_A \times N_B$ blocks transforming in the representation $R_A \otimes R_B$.
Explicitly, writing $A=(0,I)$ with $I=1,2,3$, we have
\begin{equation}
R_0\otimes R_A = R_A \quad\mbox{and}\quad R_I\otimes R_J = \delta_{IJ}  R_0 + |\epsilon_{IJK}|
\,R_K~.
\label{RR}
\end{equation}
The invariant components of $V$ and $\Phi^I$ surviving the orbifold projection are
given by those blocks 
where the non-trivial transformation properties of the fields are compensated by those of
their Chan-Paton indices and are
\begin{equation}
\big\{ V\big\}^{u_A}_{~v_A} ~\cup ~\big\{\Phi^I\big\}^{u_A}_{~v_{A\otimes I}}~.
\label{gauge}
\end{equation}
Here the symbol $\{\,\}^{u_A}_{~ v_B}$ denotes the components of the $N_A \times N_B$ block, and 
the subindex $A\otimes I$ is a shorthand for the representation product $R_A \otimes R_I$, namely
\begin{equation}
 0\otimes I = I\quad\mbox{and}\quad J\otimes I = |\epsilon_{JIK}| \,K
\label{otimes}
\end{equation}
as follows from (\ref{RR}). 
Eq. (\ref{gauge}) represents the field content of a ${\mathcal N}=1$
gauge theory with gauge group $\prod_A \mathrm U(N_A)$
and matter in the bifundamentals
$\big({N_A},{\overline N_B}\big)$, which is encoded in
the quiver diagram displayed in Fig. \ref{fig:quiver}.

The projected theory is invariant under the $\mathrm{U}(1)^3$ global symmetries
corresponding to the Cartan subgroup of the $\mathrm{SO}(6)$ ${\cal R}$-symmetry 
invariance of the ${\mathcal N}=4$ action (\ref{actgauge0}):
 \begin{equation}
\begin{aligned}
&\Phi^I\to \ee^{\ii\zeta_I}\, \Phi^I\quad,\quad V\to V\quad,\quad
W_\alpha \to \ee^{\frac{\ii}{2}\sum_I \zeta_I} \,W_\alpha\\
& d\theta\to  \ee^{-\frac{\ii}{2}\sum_I \zeta_I} \, d\theta\quad,
\quad d\bar \theta\to  \ee^{\frac{\ii}{2}\sum_I \zeta_I}\, d\bar \theta 
 ~;\end{aligned} 
\label{symgauge}
\end{equation}
these transformations encode the charges $q_I$  of the various fields w.r.t. to
the three $\mathrm{U}(1)$'s. The symmetry extends, 
as we will see, to the zero modes of the gauge fields in instantonic sectors, and can be exploited to constrain the form of the allowed non-perturbative interactions.
To this aim, it will prove useful to take linear combinations of the $\mathrm{U}(1)^3$ symmetries (\ref{symgauge}) corresponding to
introducing the charges
\begin{equation}
\label{new_charges}
q= q_1 + q_2 + q_3~, \quad\quad 
q'= q_1 - q_2 ~, \quad\quad 
q''= q_1 - q_3~. 
\end{equation}
The values of these charges for the various gauge fields are displayed in Table \ref{tableQG}.
 \begin{table}[ht]
\centering
\begin{tabular}{c|cccc}
\hline\hline
     fields$\phantom{\vdots}$  & $A_\mu$ &$\Lambda^\alpha$ 
     & $\phi^I$& $\psi^{\alpha I}$ \\
  \hline
charge $\phantom{\vdots}q$  &$0$ & $+\frac32$   & $+1$ & $-\frac12$  \\
charge $\phantom{\vdots}q^\prime$  &$0$ &  $0$
 & $\delta_{I1}-\delta_{I2}$ & $\delta_{I1}-\delta_{I2}$\\
charge $\phantom{\vdots}q^{\prime\prime}$  &$0$ &$0$
 &$\delta_{I1}-\delta_{I3}$  & $\delta_{I1}-\delta_{I3}$   \\
\hline\hline
\end{tabular}
\caption{The charges $q$, $q^\prime$ and $q^{\prime\prime}$ of the various fields of the
${\mathcal N}=1$ quiver gauge theory.  Here $A_\mu$ is the gauge vector, $\Lambda^\alpha$ is the corresponding gaugino while $\psi^{\alpha I}$ is the fermion of the
bifundamental matter superfield $\Phi^I$ of which $\phi^I$ is the
lowest component. The complex conjugate fields $\bar \phi_I$,
$\bar \Lambda_{\dot\alpha}$ and $\bar\psi_{\dot\alpha I}$ transform oppositely to the ones displayed. Finally $\theta$ transforms as $\Lambda$ and oppositely to $\bar \theta$.}
\label{tableQG}
\end{table}

The fractional D3-branes can also be thought of as D5-branes
wrapping exceptional ({\it i.e.} vanishing) 2-cycles $\mathcal C_A$ of the orbifold.
Note that there are
only three independent such cycles on $\mathbb{C}^3/\big(\mathbb{Z}_2\times \mathbb{Z}_2\big)$
which are associated to the three exceptional $\mathbb{P}^1$'s corresponding to the non-trivial elements $h_I$ of $\mathbb{Z}_2\times \mathbb{Z}_2$.
This implies that only three linear combinations of the $\mathcal C_A$'s are really independent.
Indeed, the linear combination $\sum_{A=0}^3\mathcal C_A$ is trivial in the homological sense since
a D5 brane wrapping this cycle transforms in the regular representation and can
move freely away from the singularity because it is made of a D5-brane plus
its three images under the orbifold group. The gauge kinetic functions
$\tau_A$ of the four $\mathrm{U}(N_A)$ factors can be expressed in terms of the three
K\"ahler parameters describing the complexified string volumes of the three non-trivial 
independent 2-cycles and the axion-dilaton field $\tau$.
In the unresolved (singular) orbifold limit, which from the string point of view corresponds
to switching off the fluctuations of all twisted closed string fields, we simply have
\cite{Bertolini:2001gg,Billo:2008sp}
\begin{equation}
\tau_A = \frac{\theta_A}{2\pi} + \ii\,\frac{4\pi^2}{g_A^2} = \frac{1}{4}\,\tau
\label{tauA}
\end{equation}
for all $A$'s. However, by turning on twisted closed string moduli, one can introduce
differences among the $\tau_A$'s and thus distinguish the gauge couplings of
the various group factors. 

\subsection{The instanton moduli space}
\label{subsec:moduli}
A very similar analysis applies to instantonic sectors of the gauge theory.
In this framework instantons are realized with D$(-1)$-branes and their moduli space
is described by the lowest modes of open strings with at least one end-point on the
D$(-1)$-branes. The $\mathrm U(N)$ gauge theory dynamics in the sector with
instanton number $k$ can be efficiently described
by the $\alpha' \to 0$ limit of the action of a system of $k$ D$(-1)$-branes and $N$ D3-branes
on $\mathbb{C}^3$.
This system is described in terms of
a $\mathrm{U}(N) \times \mathrm{U}(k)$ matrix theory whose action is \cite{Dorey:2002ik}
\begin{equation}
S_{\mathrm{D3/D(-1)}}={\mathrm{Tr}}_{k}\,\left[\frac{1}{2g_0^2}\,
S_G+S_{K}+S_{D}+S_\phi\right]
\label{cometipare}
\end{equation}
with
\begin{equation}
\begin{aligned}
S_{G} =& ~ D_{c}D^{c}-\frac{1}{2}\,[\chi_m,\chi_n]^2 - {\ii}\,
{\lambda}_{\dot{\alpha}A}\,[{\chi}^{AB},{\lambda}^{\dot{\alpha}}_{~B}] ~,\\
S_{K} =& ~ \chi_m \bar{w}_{\dot\alpha} w^{\dot\alpha}\chi^m -[\chi_m,a_{\mu}]^2
-\ii\,M^{\alpha A}\,[\overline\chi_{AB},M_{\alpha}^{~B}]+ \frac{\ii}{2}\,
\overline\chi_{AB} \,\bar{\mu}^{A} \mu^{B} ~,\\
S_{D} =& ~ \ii\,D_{c}\big(\bar{w}_{\dot\alpha}
(\tau^c)^{\dot\alpha}_{~\dot\beta} w^{\dot\beta} -\ii\,
\bar{\eta}_{\mu\nu}^c [a^\mu,a^\nu] \big)
+\ii \,\lambda_{\dot\alpha A}
\big( \bar{\mu}^{A} w^{\dot{\alpha}}
+\bar{w}^{\dot{\alpha}}\mu^{A} -[a^{\alpha\dot{\alpha}},M_{\alpha}^{~A}]\big)
 ~,\\
S_\phi = &\frac{1}{8}\,\epsilon^{ABCD}\bar w_{\dot\alpha}\overline\phi_{AB}
\overline\phi_{CD}w^{\dot\alpha}+\frac{1}{2}\,\bar w_{\dot\alpha}\phi^{AB}
w^{\dot\alpha}\overline\chi_{AB}
+\frac{\ii}{2}\,\bar\mu^A\overline\phi_{AB}\mu^B~.
\end{aligned}
\label{Sd1d3}
\end{equation}
The fields entering in (\ref{Sd1d3}) represent the lowest modes of open strings with
D$(-1)$/D$(-1)$ and D$(-1)$/D3 boundary conditions and are
\begin{equation}
\mathfrak M= \big\{ a_{\mu}, \chi_m, D_c, M^{\alpha A},
\lambda_{\dot\alpha A}\big\}^i_{~j} ~\cup~
\big\{ w_{\dot \alpha}, \mu^A \big\}^u_{~i} ~\cup~ \big\{
\bar w_{\dot \alpha}, \bar\mu^A \big\}^i_{~u}
\label{mod40}
\end{equation}
with $i,j=1,\ldots,k $ and $u,v=1,\ldots,N$ labeling the $k$ D$(-1)$ and the $N$ D3 boundaries respectively.
The other indices run over the following domains:
$\mu,\nu=0,\ldots,3$; $\alpha,\dot\alpha=1,2$; $m,n=1,\ldots,6$; $A,B=0,\ldots,3$, labeling,
respectively, the vector and spinor representations of the $\mathrm{SO}(4)$ Lorentz group and of
the $\mathrm{SO}(6) \sim \mathrm{SU}(4)$ internal-symmetry group%
\footnote{See footnote \ref{foot:1}.}, while $c=1,2,3$.
We have also defined
\begin{equation}
\begin{aligned}
\chi^{AB} &= \chi_m \,(\Sigma^{m})^{AB}\quad,\quad
\overline{\chi}_{AB}= \chi_m\,(\overline\Sigma^{\,m})_{AB}
= \frac{1}{2}\epsilon_{ABCD}\,\chi^{CD}~,\\
\phi^{AB} &= \phi_m \,(\Sigma^{m})^{AB}\quad,\quad
\overline{\phi}_{AB}= \phi_m\,(\overline\Sigma^{\,m})_{AB}
= \frac{1}{2}\epsilon_{ABCD}\,\phi^{CD}
\end{aligned}
\label{chi}
\end{equation}
where $\Sigma^m$ and $\overline{\Sigma}^{\,m}$ are the chiral and anti-chiral blocks of the
Dirac matrices in the six-dimensional internal space, and $\phi_m$ are the six 
vacuum expectation values of the scalar fields $\Phi^I$ in the real basis.
Finally,
\begin{equation}
\frac{1}{g^2_0}=\frac{\pi}{g_s}\,(2\pi \alpha')^{2}
\label{g0}
\end{equation}
is the coupling constant of the gauge theory on the D$(-1)$ branes.
The scaling dimensions of
the various moduli appearing in (\ref{Sd1d3}) are listed in Tab. \ref{dimensions}.
\begin{table}[ht]
\centering
\begin{tabular}{c|ccccccccc}
\hline\hline
  moduli & $\phantom{\vdots}a_\mu$ & $\chi_m$ & $D_c$  & $M^{\alpha A}$ & $\lambda_{\dot\alpha A}$
& $w_{\dot\alpha}$ & $\bar w_{\dot\alpha}$ & $\mu^A$ & $\bar \mu^A$ \\
    \hline
  dimension & $\phantom{\vdots} M_s^{-1}$ & $M_s$ & $M_s^2$ & $M_s^{-1/2}$ & $M_s^{3/2}$
& $M_s^{-1}$ & $M_s^{-1}$ & $M_s^{-1/2}$ & $M_s^{-1/2}$ \\
   \hline\hline
 \end{tabular}
 \caption{Scaling dimensions of the ADHM moduli in terms of the string
scale $M_s=(2\pi\alpha')^{-1/2}$.}
 \label{dimensions}
 \end{table}

The action (\ref{cometipare}) follows from dimensional reduction of the six-dimensional
action of the D5/D9 brane system down to zero dimensions and, as discussed in detail in Ref. \cite{Billo:2002hm}, it can be explicitly derived from scattering amplitudes of open strings with
D$(-1)$/D$(-1)$ or D$(-1)$/D3 boundary conditions on (mixed) disks.
In the field theory limit $\alpha'\to 0$ ({\it {i.e.}} $g_0\to \infty$), the term $S_G$ 
in (\ref{cometipare}) can be discarded and the fields $D^c$ and $\lambda_{\dot \alpha A}$
become Lagrange multipliers for the super ADHM constraints that realize 
the D- and F-flatness conditions in the matrix theory.

Now let us consider the $\mathbb Z_2 \times \mathbb Z_2$ orbifold projection. The
group $\mathrm{U}(N)\times \mathrm{U}(k)$ breaks down
to $\prod_A \mathrm{U}(N_A)\times \mathrm{U}(k_A)$ with $N_A$ and $k_A$ being the numbers
of fractional D3 and D$(-1)$ branes of type $A$ such that
\begin{equation}
N=\sum_{A=0}^3 N_A \quad\mbox{and}\quad k=\sum_{A=0}^3 k_A~.
\label{nk}
\end{equation}
Consequently, the indices $u$ and
$i$ break into $u_A=1,\ldots,N_A$ and $i_A=1,\ldots,k_A$,
while the spinor index $A$ splits into $A=(0,I)$ with $A=0$ denoting the ${\mathcal N}=1$
unbroken symmetry. For the sake of simplicity from now on we always omit the index 0
and write
\begin{equation}
M^{\alpha 0}\equiv M^{\alpha}\quad,\quad
\lambda_{\dot\alpha 0}\equiv \lambda_{\dot\alpha} \quad,\quad
\mu^0\equiv \mu \quad,\quad\bar\mu^0\equiv\bar\mu~.
\label{00}
\end{equation}
Furthermore we set
\begin{equation}
\begin{aligned}
\overline{\chi}_{0I} &\equiv \overline{\chi}_{I}\quad,&\quad  &\chi^{0I} =\frac{1}{2}\,\epsilon^{IJK}\overline{\chi}_{JK}\equiv\chi^I~,\\
\overline{\phi}_{0I} &\equiv \overline{\phi}_{I}=\langle \,\bar{\Phi}_I\rangle
\quad,&\quad  &\phi^{0I}=\frac{1}{2}\,\epsilon^{IJK}\overline{\phi}_{JK}\equiv \phi^I
=\langle \,{\Phi}^I\rangle~.
\label{varchi}
\end{aligned}
\end{equation}
This notation makes more manifest which zero-modes couple to the holomorphic superfields
and which others couple to the anti-holomorphic ones. Indeed, 
the action $S_\phi$ in (\ref{Sd1d3}) becomes
\begin{equation}
\begin{aligned}
S_\phi = &~\frac{1}{2}\,\bar{w}_{\dot\alpha}\big(\phi^I\bar\phi_I+
\bar\phi_I\phi^I\big) w^{\dot\alpha} + \bar w_{\dot\alpha} \phi^I 
 w^{\dot\alpha} \overline\chi_{I} + \chi^{I} \bar w_{\dot\alpha}
 \bar\phi_I w^{\dot\alpha}\\
 &-\frac{\ii}{2}\,\bar\mu \,\bar\phi_I \mu^I
+\frac{\ii}{2}\,\bar\mu^I \bar\phi_I \mu  -\frac{\ii}{2}\,
 \epsilon_{IJK}\bar\mu^I\phi^J\mu^K~.
\end{aligned}
\label{Sphi1}
\end{equation}

Taking into account the $\mathbb Z_2 \times \mathbb Z_2$ transformation properties 
of the various fields and of their Chan-Paton labels, one finds that the moduli that survive
the orbifold projection are
\begin{equation}
\begin{aligned}
{\mathfrak M} ~=~& \big\{ a_{\mu}, D_c, M^{\alpha}, \lambda_{\dot \alpha}\big\}^{i_A}_{~ j_A}
     ~~\cup ~~ \big\{ w_{\dot \alpha},   \mu \big\}^{u_A}_{~j_A}
~~ \cup ~~\big\{\bar w_{\dot \alpha},  \bar\mu \big\}^{i_A}_{~ u_A}
\\
&\cup~~\big\{\chi^I, \overline{\chi}_{I}, M^{\alpha I},
\lambda_{\dot\alpha I}\big\}^{i_A}_{~ j_{A\otimes I}}
~~\cup~~\big\{\mu^I\big\}^{u_A}_{~ i_{A\otimes I}}
~~\cup~~\big\{\bar\mu^I\big\}^{i_A}_{~ u_{A\otimes I}}~.
\end{aligned}
\label{mod4}
\end{equation}
Like for the chiral multiplets in (\ref{gauge}), the non-trivial transformation
of the instanton moduli carrying an index $I$ is compensated by a similar transformation 
of the Chan-Paton labels making the whole expression invariant under the orbifold
group, as indicated in the second line of (\ref{mod4}).

The $(\mathbb Z_2 \times \mathbb Z_2)$ projected moduli action is invariant under the $\mathrm U(1)^3\subset \mathrm{SO}(6)$ symmetry
whose properties on the D3/D3 sector we already discussed in Sec. \ref{subsec:gauge}, see Table \ref{tableQG}.
The charges of the moduli with respect to the same choice of $\mathrm{U}(1)^3$  made in Eq. (\ref{new_charges})
are given in Table \ref{tableu1}.
\begin{table}[ht]
\centering
\begin{tabular}{c|cccccc}
\hline\hline
     moduli$\phantom{\vdots}$  & $\chi^I$ & $\bar \chi_I$ &
     $\mu,\bar \mu,M^\alpha$ & $\mu^I,\bar \mu^I,M^{\alpha I}$  & $\lambda_{\dot\alpha}$
 &$\lambda_{\dot\alpha I}$ \\
  \hline
charge $\phantom{\vdots}q$  & $+1$ & $-1$ & $+\frac32$ & $-\frac12$ & $-\frac32$ & $+\frac12$ \\
charge $\phantom{\vdots}q^\prime$   &  $\delta_{I1}-\delta_{I2}$
&  $\delta_{I2}-\delta_{I1}$  & $0$ & $\delta_{I1}-\delta_{I2}$
&  $0$ & $\delta_{I2}-\delta_{I1}$\\
charge $\phantom{\vdots}q^{\prime\prime}$ &$\delta_{I1}-\delta_{I3}$
& $\delta_{I3}-\delta_{I1} $& $0$ & $\delta_{I1}-\delta_{I3}$ &
$0$ & $\delta_{I3}-\delta_{I1}$ \\
\hline\hline
\end{tabular}
\caption{The charges $q$, $q^\prime$ and $q^{\prime\prime}$ of the various fields of the
D3/D$(-1)$ brane system. The bosonic moduli  $a^\mu$, $w_{\dot\alpha}$,
$\bar w_{\dot\alpha}$ and $D_c$ are neutral under all three $\mathrm U(1)$'s.}
\label{tableu1}
\end{table}

Among the moduli in ${\mathfrak M}$, the bosonic combinations
\begin{equation}
x^\mu \equiv \frac{1}{k}\,\sum_{A=0}^3\,\sum_{{i_A}=1}^{k_A} \big\{a^\mu\big\}^{i_A}_{~i_A}
\label{xmu}
\end{equation}
represent the center of mass coordinates of the D$(-1)$-branes and can be interpreted as the
Goldstone modes associated to the translational symmetry of the D3-branes that is broken by the
D-instantons. Thus they can be identified with the space-time coordinates, and indeed have dimensions
of a length. Similarly, the fermionic combinations
\begin{equation}
\theta^\alpha \equiv \frac{1}{k}\,\sum_{A=0}^3\,\sum_{{i_A}=1}^{k_A} \big\{M^\alpha\big\}^{i_A}_{~i_A}
\label{theta}
\end{equation}
are the Goldstinos for the two supersymmetries of the D3-branes that are broken
by the D$(-1)$-branes, and thus they can be identified with the chiral fermionic superspace coordinates.
Indeed they have dimensions of (length)$^{1/2}$.
Notice that neither $x^\mu$ nor $\theta^\alpha$ appear in the moduli action obtained by
projecting (\ref{cometipare}).

The moduli (\ref{mod4}) account for both gauge and stringy instantons. 
Gauge instantons correspond to D$(-1)$-branes that sit on non-empty
nodes of the quiver diagram, so that their number $k_A$ can be interpreted as the second Chern class
of the Yang-Mills bundle of the $\mathrm{U}(N_A)$ component of the gauge group.
Stringy instantons correspond instead to D$(-1)$-branes
occupying empty nodes of the quiver,
so that in this case we can set $k_A N_A=0$ for all $A$'s.
{F}rom (\ref{mod4}) 
one sees that in the stringy instanton case
the modes $w_{\dot \alpha}$, $\bar w_{\dot \alpha}$, $\mu$ and $\bar\mu$ 
are missing since there are no $k_A \times N_A$ invariant blocks and thus
the only moduli are the charged fermionic fields $\mu^I$ and $\bar\mu^I$.
Recalling that the bosonic $w$-moduli describe the instanton sizes and gauge orientations, one can
say that exotic instantons are entirely specified by their spacetime positions.
The presence (absence) of bosonic modes in the D3/D$(-1)$ sector for gauge (stringy)
instantons can be understood in geometric terms by blowing up 2-cycles at the singularity
and reinterpreting the fractional D3/D$(-1)$ system in terms of a D5/E1 bound state wrapping
an exceptional 2-cycle of $\mathbb{C}^3/\big(\mathbb{Z}_2\times \mathbb{Z}_2\big)$.
Gauge (stringy) instantons correspond to the cases when the D5 brane and the E1 are (are not)
parallel in the internal space. In the first case the number of Neumann-Dirichlet
directions is 4 and therefore the NS ground states is massless. In the stringy case, the number
of Neumann-Dirichlet directions exceeds 4 and
therefore the NS ground state is massive and all charged moduli come only from the fermionic R sector.

\section{D-instanton partition function}
\label{sec:effint}

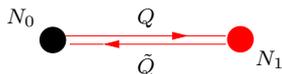
\begin{figure}[hbt]
\begin{center}
\begin{picture}(0,0)%
\includegraphics{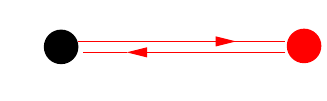}%
\end{picture}%
\setlength{\unitlength}{1381sp}%
\begingroup\makeatletter\ifx\SetFigFontNFSS\undefined%
\gdef\SetFigFontNFSS#1#2#3#4#5{%
  \reset@font\fontsize{#1}{#2pt}%
  \fontfamily{#3}\fontseries{#4}\fontshape{#5}%
  \selectfont}%
\fi\endgroup%
\begin{picture}(4455,1375)(361,-1097)
\put(2701,-61){\makebox(0,0)[lb]{\smash{{\SetFigFontNFSS{8}{9.6}{\familydefault}{\mddefault}{\updefault}$Q$}}}}
\put(2701,-961){\makebox(0,0)[lb]{\smash{{\SetFigFontNFSS{8}{9.6}{\familydefault}{\mddefault}{\updefault}${\tilde Q}$}}}}
\put(4801,-811){\makebox(0,0)[lb]{\smash{{\SetFigFontNFSS{8}{9.6}{\familydefault}{\mddefault}{\updefault}$N_1$}}}}
\put(376,-61){\makebox(0,0)[lb]{\smash{{\SetFigFontNFSS{8}{9.6}{\familydefault}{\mddefault}{\updefault}$N_0$}}}}
\end{picture}%
\end{center}
\caption{This simple quiver gauge theory is (from the point of view of one of the nodes) 
just $\mathcal{N}=1$ SQCD.}
\label{fig:quiver_SQCD}
\end{figure}

In this section we study the non-perturbative effects generated by fractional
D-instantons on the $\mathcal N=1$ gauge theory realized with space-filling fractional D3 branes.
For simplicity from now on we take (see Fig. \ref{fig:quiver_SQCD})
\begin{equation}
N_2=N_3=0\quad\mbox{with}\quad N_0~\mbox{and}~N_1~\mbox{arbitrary}~,
\label{NA}
\end{equation}
which is the simplest configuration that allows us to discuss both gauge and
stringy instanton effects.

This brane system describes a $\mathcal N=1$ theory with gauge group 
$\mathrm U(N_0)\times \mathrm U(N_1)$ and a single bifundamental multiplet
\begin{equation}
 \label{Phimult}
\Phi^1(x,\theta)\equiv \Phi(x,\theta)=\phi(x)+\sqrt2 \theta \psi(x)+ \theta^2 F(x)~,
\end{equation}
which in block form is
\begin{equation}
\Phi=\left(
         \begin{array}{cc}
           0 & Q^{u}_{~ f} \\
           \widetilde Q^{f}_{~ u} & 0 \\
         \end{array}
       \right)
\label{Phi}
\end{equation}
with $u=1,\ldots N_0$ and $f=1,\ldots N_1$. The two off-diagonal blocks $Q$ and $\widetilde Q$ represent the quark and anti-quark superfields
which transform respectively in the fundamental and anti-fundamental of $\mathrm U(N_0)$, and in the anti-fundamental and fundamental of $\mathrm U(N_1)$. Both quarks and anti-quarks are neutral
under the diagonal $\mathrm U(1)$ factor of the gauge group, which decouples. 
On the other hand \cite{Intriligator:2005aw}
the relative $\mathrm U(1)$ group, under which both $Q$ and $\widetilde Q$ are charged, is IR free 
and thus at low energies the resulting effective gauge group
is $\mathrm {SU}(N_0)\times \mathrm {SU}(N_1)$. Therefore,
from the point of view of, say, the $\mathrm{SU}(N_0)$ factor
this theory is just $\mathcal N=1$ SQCD with $N_c=N_0$ colors
and $N_f=N_1$ flavors. In the following we will study
the non-perturbative properties of this theory in the
Higgs phase where the gauge invariance is completely broken by
giving (large) vacuum expectation values to the lowest components of the matter superfields.
This requires $N_f\geq N_c-1$. The moduli space of this SQCD is obtained by imposing
the D-flatness conditions. As remarked in \cite{Franco:2005zu}, even if the effective gauge
group is $\mathrm{SU}(N_c)$, we have to impose the D-term equations also for the
(massive) \footnote{For a discussion of this point in the same orbifold model we are
considering, see for example the beginning of Section 3 of the published version of Ref. \cite{Matsuo:2008nu}.}
$\mathrm{U}(1)$ factors to obtain the correct moduli space of the quiver theory; 
in our case these D-term conditions lead to the constraint
\begin{equation}
Q\,\bar Q - \bar{\widetilde Q}\,\widetilde Q = \xi\,\one_{N_c\times N_c}
\label{D-flat1}
\end{equation}
where $\xi$ is a Fayet-Iliopoulos parameter related to twisted closed string fields which vanish
in the singular orbifold limit. For $N_f\geq N_c$ the D-term constraints allow for flat directions parameterized by meson fields
\begin{equation}
M^{f_1}_{~f_2} \equiv {\widetilde Q}^{f_1}_{~u} \,Q^{u}_{~f_2}
\label{meson}
\end{equation}
and baryon fields
\begin{equation}
B_{f_1\ldots f_{N_c}}=\epsilon_{u_1\ldots u_{N_c}}\, Q^{u_1}_{~f1} \ldots Q^{u_{N_c}}_{~\,f_{N_c}} 
\quad,\quad
{\widetilde B}^{f_1\ldots f_{N_c}}
=\epsilon^{u_1\ldots u_{N_c}}\, {\widetilde Q}^{f1}_{~u_1} \ldots 
{\widetilde Q}^{f_{N_c}}_{~\,u_{N_c}} 
\label{baryon}
\end{equation}
which are subject to constraints whose specific form depends on the difference $(N_f-N_c)$ (see for instance Ref. \cite{Intriligator:1995au}). These are the good observables 
of the low-energy theory in the Higgs phase.
For $N_f=N_c-1$, instead, the baryons cannot be formed and only the meson fields are present.

To have a quick understanding of the non-perturbative effects that
can be obtained in our stringy set-up, it is convenient to use dimensional analysis and exploit the
symmetries of the D3/D$(-1)$ brane system; we will see that besides the well-known
one-instanton effects like the ADS superpotential at $N_f=N_c-1$ \cite{Affleck:1983mk}, in the
quiver theory an infinite tower of multi-instanton corrections to the superpotential are 
in principle allowed. This is what we are going to show in the remainder of this section.
In Section \ref{secn:1inst} we specialize our discussion to the one-instanton sector and,
using again dimensional analysis and symmetry considerations, we analyze various types of non-perturbative effects in the low-energy theory, 
as a preparation for the study of the flux-induced terms presented 
in Sections \ref{secn:fluxeffects} and \ref{secn:strinst}.

\subsection{The moduli space integral}
\label{integral}
The non-perturbative effects produced by a configuration of fractional D-instantons 
with numbers $k_A$ can be analyzed by studying the centered partition function
\begin{equation}
{W}_{\mathrm{n.p.}}  = \int d\,{\widehat{\mathfrak M}}\,\,\,\prod_{A=0}^3\!\Big(
M_s^{\,k_A \beta_A}\,\ee^{2\pi \ii k_A \tau_A }\,
\ee^{-\mathrm{Tr}_{k_A}[S_K+S_D+S_\phi]}\Big)~,
\label{Z}
\end{equation}
where the integration is over all moduli listed in (\ref{mod4}) except for the
center of mass supercoordinates $x^\mu$ and $\theta^\alpha$ defined in (\ref{xmu}) and
(\ref{theta}). These centered moduli are collectively denoted by $\widehat{\mathfrak M}$.
The action $\mathrm{Tr}_{k_A}[S_K+S_D+S_\phi]$ is obtained by taking the field theory limit in
(\ref{cometipare}) and restricting
the moduli to their $\mathbb Z_2 \times \mathbb Z_2$ invariant blocks for each $A$, while
the term $2\pi\ii k_A\tau_A$ represents the classical action of $k_A$ fractional
D-instantons of type $A$ (see Eqs. (\ref{wfluxnp}) and (\ref{tauA})).
Finally the power of the string scale $M_s$ compensates for the scaling dimensions
of the measure over the centered moduli space so that
the centered partition function ${W}_{\mathrm{n.p.}}$ has mass dimension 3,
as expected.
Indeed, using Tab. \ref{dimensions} one can easily show that the mass dimension $D$
of the instanton measure is
\begin{equation}
D\big[d\,{\widehat{\mathfrak M}}\big]= 3 -\sum_{A=0}^3 k_A\beta_A
\label{DdM}
\end{equation}
where $\beta_A$ 
is the one-loop $\beta$-function coefficient of the $\mathcal N=1$ $\mathrm{SU}(N_A)$ gauge theory
with $\sum_{I=1}^3 N_{A\otimes I}$ fundamentals and anti-fundamentals, namely
\begin{equation}
\beta_A = 3\, \ell(\mathrm{Adj}_A)- 2\sum_{I=1}^3 
N_{A \otimes I}\,\,\ell(N_A)
=3 N_A-\sum_{I=1}^3 N_{A\otimes I}
 \label{betagauge}
 \end{equation} 
where $\ell({r})$ denotes the index%
\footnote{The index $\ell(r)$ is defined by 
${\mathrm Tr}_{r}\big(T^a T^b\big)=\ell(r) \,\delta^{ab}$. 
For $\mathrm{SU}(N)$ gauge groups, the indices in the adjoint, fundamental, symmetric and 
antisymmetric representations are, respectively, given by  $\ell(\mathrm{Adj})=N$, $\ell(N)=\frac12$,
$\ell\big(\frac12N(N+1)\big)=N+2$ and $\ell\big(\frac12N(N-1)\big)=N-2$.}
of the representation ${r}$.
It is interesting to remark that the explicit expression of $\beta_A$
is well-defined even in the case
$N_A=0$ where it cannot be interpreted as the $\beta$-function coefficient 
of any gauge theory. Keeping this in mind, all formulas in this section can be
applied to both gauge and stringy instanton configurations.

Coming back to the centered partition function (\ref{Z}) one can ask which dependence on
the scalar vacuum expectation values is generated by the integral over the instanton
moduli. A quick answer to this question follows by requiring that the form of
${W}_{\mathrm{n.p.}}$ be consistent with the symmetries of the D3/D$(-1)$ system.
In particular we can exploit the $\mathrm U(1)^3$ symmetries left
unbroken by the orbifold projection that we have discussed in Section \ref{secn:d3d-1}.
These are symmetries of the D3/D$(-1)$ action but not of the instanton measure. Indeed, since
there are unpaired moduli, like $\mu^A$ and $\bar\mu^A$, which transform in the same way under $U(1)^3$,
the charges of the centered instanton measure, and hence of ${W}_{\mathrm{n.p.}}$, 
are non-trivial. In particular, the charge $q$ is 
\footnote{As one can see from Tab. \ref{tableu1}, the moduli $M^{\alpha I}$
and $\lambda_{\dot \alpha I}$ have opposite charges, like
$M^{\alpha}$ and $\lambda_{\dot\alpha}$. For this reason they do not
contribute to the total charges of the centered measure $d\,{\widehat{\mathfrak M}}$ which thus
depend only on the charges of unpaired moduli.}
 \begin{equation}
q\big[d\,{\widehat{\mathfrak M}}\big]=-2n_\mu q(\mu)-2n_{\mu^I} q(\mu^I)-2 q(\lambda)
\label{condq}
\end{equation}
where $n_\mu$ and $n_{\mu^I}$ are the numbers of $\mu$ and $\mu^I$ moduli, and the factors 
of $2$ account for identical contributions from $\bar\mu$ and $\bar \mu^I$. Finally,
the term $-2q(\lambda)$ comes from the two components of the anti-chiral fermion
\begin{equation}
 \lambda_{\dot\alpha}
 \equiv \frac{1}{k}\,\sum_{A=0}^3\,\sum_{{i_A}=1}^{k_A} \big\{\lambda_{\dot\alpha}\big\}^{i_A}_{~i_A}
 \label{lambda}
 \end{equation} 
which are unpaired since their partners, namely the fermionic superspace coordinates $\theta^\alpha$ defined in (\ref{theta}), have been taken out from the centered measure
$d\,{\widehat{\mathfrak M}}$. The minus signs in (\ref{condq}) come from the fact that a
fermionic differential transforms oppositely to the field itself.
Using the charges listed in Tab. \ref{tableu1}, it is easy to rewrite (\ref{condq}) as
\begin{equation}
q\big[d\,{\widehat{\mathfrak M}}\big]= 3-\sum_{A=0}^3k_A\Big(3
N_A-\sum_{I=1}^3 N_{A\otimes I}\Big)=3-\sum_{A=0}^3 k_A \beta_A~.
\label{qdm}
\end{equation}
In a similar way one finds
\begin{equation}
\begin{aligned}
q^\prime\big[d\,{\widehat{\mathfrak M}}\big] &=-2n_\mu q^\prime(\mu)
-2n_{\mu^I} q^\prime(\mu^I)-2 q^\prime(\lambda)=-2 \sum_{A=0}^3k_A \big(N_{A\otimes 1}-
N_{A\otimes 2}\big)
~,\\
q^{\prime\prime}\big[d\,{\widehat{\mathfrak M}}\big] &= -2n_\mu q^{\prime\prime}(\mu)-2n_{\mu^I} q^{\prime\prime}(\mu^I)-2 q^{\prime\prime}(\lambda)=-2 \sum_{A=0}^3k_A \big(N_{A\otimes 1}
-N_{A\otimes 3}\big)~.
\end{aligned}
\label{qdm1}
\end{equation}
One can check that the $\mathrm{U}(1)^3$ charges of the ADHM measure coincide
with the ones of the moduli space of instanton zero-modes. In fact, since in an instanton 
background the bosonic zero-modes always come together with their complex conjugates, 
the charges $q$, $q^\prime$ and $q^{\prime\prime}$ of the instanton measure depend
only on the number of fermionic zero-modes, namely on the number $n_{\Lambda}$ of gaugino zero-modes, 
and on the number $n_{\psi^I}$ of zero-modes of the fundamental matter fields%
\footnote{Remember that in an instanton background $\bar\Lambda=\bar\psi_I=0$}.
These numbers are given by the index of the Dirac operator evaluated,
respectively, in the adjoint and fundamental $\mathrm{SU}(N_A)$ representations 
under which the fields transform, {\it i.e.}
\begin{equation}
\begin{aligned}
n_{\Lambda} &=2 k_A \,\ell(\mathrm{Adj}_A)=2 k_A \,N_A~,\\
n_{\psi^I} &= 2 k_A\,\left(2 N_{A\otimes I}\right)\,\ell(N_A)=2 k_A \,N_{A\otimes I}~.
\label{indexgauge}
 \end{aligned}
\end{equation}
Taking into account the contribution of the two fermionic superspace coordinates
$\theta^\alpha$ to the charges of the instanton measure and using the values reported
in Tab. \ref{tableQG}, we have
\begin{equation}
\begin{aligned}
q&= -n_{\Lambda}\,q(\Lambda) - \sum_{I=1}^3 n_{\psi^I}\,q(\psi^I) - 2q(\theta) 
= -\frac32\,n_{\Lambda}
+\frac12\,\sum_{I=1}^3n_{\psi^I}+3~,\\
q^\prime&= -n_{\Lambda}\,q^\prime(\Lambda) - \sum_{I=1}^3 n_{\psi^I}\,q^\prime(\psi^I) - 2q^\prime(\theta)=
-n_{\psi^1}+n_{\psi^2}~,\\
q^{\prime\prime}&= - n_{\Lambda}\,q^{\prime\prime}(\Lambda) - \sum_{I=1}^3 n_{\psi^I}\,q(\psi^I) 
- 2q^{\prime\prime}(\theta)= -n_{\psi^1}+n_{\psi^3}~.
 \end{aligned}
\label{charges00}
\end{equation}
Exploiting (\ref{indexgauge}), it is immediate to see that these charges coincide with the
ones given in (\ref{qdm}) and (\ref{qdm1}) and computed using the ADHM construction.

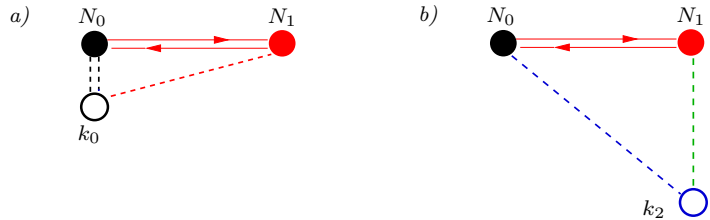
\begin{figure}[hbt]
\begin{center}
\begin{picture}(0,0)%
\includegraphics{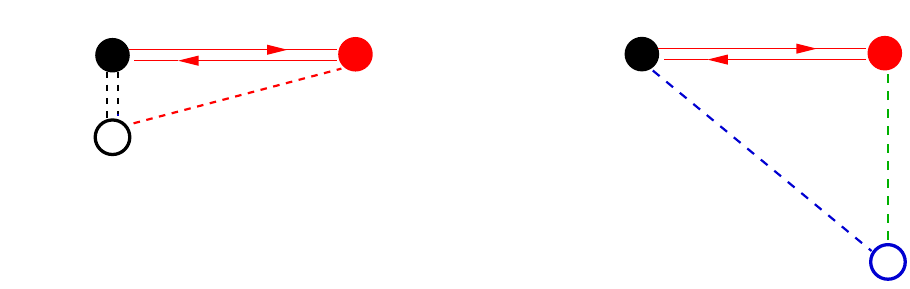}%
\end{picture}%
\setlength{\unitlength}{1381sp}%
\begingroup\makeatletter\ifx\SetFigFontNFSS\undefined%
\gdef\SetFigFontNFSS#1#2#3#4#5{%
  \reset@font\fontsize{#1}{#2pt}%
  \fontfamily{#3}\fontseries{#4}\fontshape{#5}%
  \selectfont}%
\fi\endgroup%
\begin{picture}(12447,3931)(-344,-3557)
\put(6946, 59){\makebox(0,0)[lb]{\smash{{\SetFigFontNFSS{8}{9.6}{\familydefault}{\mddefault}{\updefault}\emph{b)}}}}}
\put(901,-2086){\makebox(0,0)[lb]{\smash{{\SetFigFontNFSS{8}{9.6}{\familydefault}{\mddefault}{\updefault}$k_0$}}}}
\put(-329, 29){\makebox(0,0)[lb]{\smash{{\SetFigFontNFSS{8}{9.6}{\familydefault}{\mddefault}{\updefault}\emph{a)}}}}}
\put(901, 14){\makebox(0,0)[lb]{\smash{{\SetFigFontNFSS{8}{9.6}{\familydefault}{\mddefault}{\updefault}$N_0$}}}}
\put(4276, 14){\makebox(0,0)[lb]{\smash{{\SetFigFontNFSS{8}{9.6}{\familydefault}{\mddefault}{\updefault}$N_1$}}}}
\put(10936,-3421){\makebox(0,0)[lb]{\smash{{\SetFigFontNFSS{8}{9.6}{\familydefault}{\mddefault}{\updefault}$k_2$}}}}
\put(8161, 29){\makebox(0,0)[lb]{\smash{{\SetFigFontNFSS{8}{9.6}{\familydefault}{\mddefault}{\updefault}$N_0$}}}}
\put(11536, 29){\makebox(0,0)[lb]{\smash{{\SetFigFontNFSS{8}{9.6}{\familydefault}{\mddefault}{\updefault}$N_1$}}}}
\end{picture}%
\end{center}
\caption{D3/D$(-1)$-quiver for SQCD with \emph{a)} gauge instantons and \emph{b)} stringy instantons. 
Filled and empty circles represent stacks of D3 and D$(-1)$ branes,
solid lines stand for chiral bifundamental matter, dashed lines for charged instanton moduli.
A single dashed line represents the fermions $\mu$, while a double dashed line is a $(\mu,w)$ pair. }
\label{fk0k2}
\end{figure}

So far we have considered a generic D-instanton configuration. From now on we will focus on 
two cases, namely gauge and stringy instantons, which correspond to
the following choices of $k_A$'s
\begin{equation}
\begin{aligned}
{\mathrm{gauge}~:~~~} & \big(k_0,k_1,0,0\big)
~,\\
{\mathrm{stringy}~:~~~} & \big(0,0,k_2,k_3\big)~.
\end{aligned}
\label{gagstr}
\end{equation}
Some gauge and stringy instanton quiver diagrams are displayed in Fig. \ref{fk0k2}.

As noticed above, for stringy instantons we have $k_A N_A=0$ and therefore the only charged
ADHM moduli that survive are $\mu^I$ and $\bar\mu^I$, while $w$,
$\bar w$, $\mu$ and $\bar\mu$ are absent. As a consequence, the
fermion $\lambda$ of Eq. (\ref{lambda}) decouples from the moduli action and in the
centered partition function of stringy instantons there is an unbalanced fermionic zero-mode
integration. Therefore, unless such zero-modes are removed, for example with an orientifold projection
\cite{Argurio:2007qk,Argurio:2007vqa,Bianchi:2007wy}, or lifted with some mechanism \cite{Petersson:2007sc,GarciaEtxebarria:2008pi}, one gets a vanishing result.
We will return to the stringy instanton configurations in Section \ref{secn:fluxeffects}
where we discuss how bulk fluxes can cure this problem. In the remaining part of this section
we instead analyze in more detail the instanton partition function for a generic configuration
of gauge instantons.

\subsection{Zero-mode counting for gauge instantons}
\label{subsec:zeromode}

For gauge instantons the centered partition function (\ref{Z}) 
is a function of $\phi$ and $\bar\phi$, {\it{i.e.}} of the vacuum
expectation values of the matter superfield (\ref{Phi}) and its conjugate, 
with scaling dimension 3.
The most general ansatz for ${W}_{\mathrm{n.p.}}$ is therefore
\begin{equation}
{W}_{\mathrm{n.p.}}=  \mathcal C
\,M_s^{(k_0\beta_0+k_1\beta_1)}\,\ee^{2\pi\ii(k_0\tau_0+k_1\tau_1)}\,\,
\bar\phi^{n}\,{\phi}^{m}
\label{Wn}
\end{equation}
with $n+m+k_0\beta_0+k_1\beta_1=3$ and $\mathcal C$ a numerical constant.
Using Tab. \ref{tableQG}, it is easy to see that the $\mathrm U(1)^3$ charges of 
this expression are all equal and given by
\begin{equation}
q\big[{W}_{\mathrm{n.p.}}\big] =
q^\prime\big[{W}_{\mathrm{n.p.}}\big] =
q^{\prime\prime}\big[{W}_{\mathrm{n.p.}}\big] =3-2n -k_0\beta_0 -k_1\beta_1~.
\label{qw}
\end{equation}
We must require that these charges match those of the centered measure,
which, as follows from (\ref{qdm}) and (\ref{qdm1}), in this case are given by
\begin{equation}
q\big[d\,{\widehat{\mathfrak M}}\big] = 3-k_0\beta_0 -k_1\beta_1\quad,\quad
q^\prime\big[d\,{\widehat{\mathfrak M}}\big] = q^{\prime\prime}\big[d\,{\widehat{\mathfrak M}}\big]
=-2k_0N_1-2k_1N_0 ~.
\label{qqq}
\end{equation}
Then we immediately find that $n=0$ and
\begin{equation}
(k_0-k_1)(N_0-N_1)=1 ~.
\end{equation}
This equation is solved by
\begin{equation}
k_1=k_0-1 \quad,\quad N_1=N_0-1~,
\label{sol1}
\end{equation}
or by
\begin{equation}
k_0=k_1-1 \quad,\quad N_0=N_1-1~.
\label{sol2}
\end{equation}
Thus the partition function generated by gauge instantons is, as expected, an holomorphic function
of $\phi$ and is given by
\begin{equation}
{W}_{\mathrm{n.p.}}=  \mathcal C
\,M_s^{(k_0\beta_0+k_1\beta_1)}\,\ee^{2\pi\ii(k_0\tau_0+k_1\tau_1)}\,\,
\phi^{(3-k_0\beta_0-k_1\beta_1)}~.
\label{Wn1}
\end{equation}
Notice that this is a formal expression in which $\phi$ stands for the vacuum expectation
values of either the quark
or anti-quark superfields $Q$ and $\widetilde Q$ defined in
(\ref{Phi}), and actually only the appropriate gauge invariant combinations of these should
appear in the final result. The solutions above with $(k_0=1,k_1=0)$ and $(k_0=0,k_1=1)$ 
reproduce the well-known ADS superpotential \cite{Affleck:1983mk} 
for $\mathrm{SU}(N_0)$ and $\mathrm{SU}(N_1)$ SQCD's
respectively. The multi-instanton corrections with $k_0,k_1 > 0$ are instead a distinct feature of
the quiver gauge theory we have engineered with the fractional D3 branes.
We conclude by observing that we could have arrived at the same results without
referring to the charges of the ADHM moduli but using instead those of the gauge field
zero-modes in the instanton background. 

It is possible to generalize the previous analysis of the instanton partition
function by including a dependence on the
entire matter fields and not only on their vacuum expectation values $\phi$ and $\bar\phi$.
This leads to a very rich structure of non-perturbative interactions that include the
holomorphic ADS superpotential when $N_f=N_c-1$ and the multi-fermion F-terms of the
Beasley-Witten (BW) type \cite{Beasley:2004ys} when $N_f\geq N_c$, plus their possible
multi-instanton extensions. In the following sections we will analyze in detail such
non-perturbative effective interactions in the one-instanton case.

\section{Effective interactions from gauge instantons}
\label{secn:1inst}
In this section we discuss the non-perturbative effective interactions induced by instantons using the explicit string construction of the ADHM moduli provided by fractional D3 and D$(-1)$ branes,
and show that in the field theory limit $\alpha'\to 0$ we recover the known 
non-perturbative F-terms, such as the ADS superpotential \cite{Affleck:1983mk} 
and the BW multi-fermion couplings \cite{Beasley:2004ys}.
{From} now on we will consider one-instanton effects in the quiver gauge theory
corresponding to a D3-brane system with $N_2=N_3=0$; this should not be regarded as a limitation of
our procedure but only a choice made for the sake of simplicity.

\subsection{The gauge instanton action}
\label{subsec:instact}
To discuss the D-instanton induced effective action on the D3 brane volume in the Higgs branch,
we first have to generalize the results of Section \ref{subsec:moduli}
and introduce in the moduli action a
dependence on the entire matter superfields and not only on their vacuum expectation values.
As discussed in detail in \cite{Green:2000ke,Billo:2002hm,Billo:2006jm}, the couplings of the matter
fields with the ADHM instanton moduli can be obtained by computing mixed disk diagrams
with insertions of vertex operators for dynamical 3/3 strings on the portion of the boundary attached to the D3-branes.
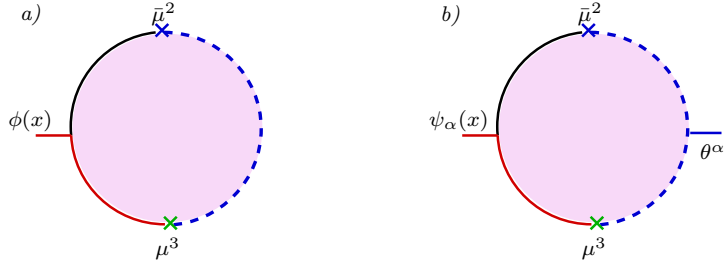
\begin{figure}[htb]
 \begin{center}
 \begin{picture}(0,0)%
\includegraphics{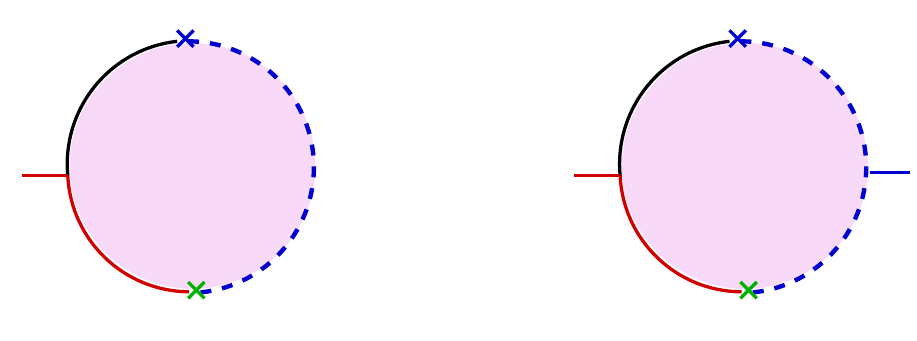}%
\end{picture}%
\setlength{\unitlength}{1381sp}%
\begingroup\makeatletter\ifx\SetFigFontNFSS\undefined%
\gdef\SetFigFontNFSS#1#2#3#4#5{%
  \reset@font\fontsize{#1}{#2pt}%
  \fontfamily{#3}\fontseries{#4}\fontshape{#5}%
  \selectfont}%
\fi\endgroup%
\begin{picture}(12513,4627)(496,-3845)
\put(311,-1426){\makebox(0,0)[lb]{\smash{{\SetFigFontNFSS{8}{9.6}{\familydefault}{\mddefault}{\updefault}$\phi(x)$}}}}
\put(2926,-3736){\makebox(0,0)[lb]{\smash{{\SetFigFontNFSS{8}{9.6}{\familydefault}{\mddefault}{\updefault}$\mu^3$}}}}
\put(2851,464){\makebox(0,0)[lb]{\smash{{\SetFigFontNFSS{8}{9.6}{\familydefault}{\mddefault}{\updefault}${\bar\mu}^2$}}}}
\put(511,479){\makebox(0,0)[lb]{\smash{{\SetFigFontNFSS{8}{9.6}{\familydefault}{\mddefault}{\updefault}\emph{a)}}}}}
\put(7981,464){\makebox(0,0)[lb]{\smash{{\SetFigFontNFSS{8}{9.6}{\familydefault}{\mddefault}{\updefault}\emph{b)}}}}}
\put(10426,464){\makebox(0,0)[lb]{\smash{{\SetFigFontNFSS{8}{9.6}{\familydefault}{\mddefault}{\updefault}${\bar\mu}^2$}}}}
\put(7786,-1426){\makebox(0,0)[lb]{\smash{{\SetFigFontNFSS{8}{9.6}{\familydefault}{\mddefault}{\updefault}$\psi_\alpha(x)$}}}}
\put(10501,-3736){\makebox(0,0)[lb]{\smash{{\SetFigFontNFSS{8}{9.6}{\familydefault}{\mddefault}{\updefault}$\mu^3$}}}}
\put(12601,-2011){\makebox(0,0)[lb]{\smash{{\SetFigFontNFSS{8}{9.6}{\familydefault}{\mddefault}{\updefault}$\theta^\alpha$}}}}
\end{picture}%
 \end{center}
\caption{Disk diagrams leading to the interaction between the scalar $\phi(x)$, or its
superpartner $\psi_\alpha(x)$, and the fermionic instanton moduli $\mu^3$ and $\bar\mu^2$.
}
\label{fig:phimu2mu3}
\end{figure}

An example of a coupling of $\phi(x)$ with the fermionic moduli $\mu$ and $\bar\mu$ is
provided by the diagram of Fig. \ref{fig:phimu2mu3}\emph{a)}, whose 
explicit evaluation leads to
\begin{equation}
 \frac{\ii}{2}\,\bar\mu^2\,\phi(x)\,\mu^3~.
\label{phimumu}
\end{equation}
If $\phi(x)$ is frozen to its vacuum expectation value, this coupling precisely accounts for one
of the last terms of $S_\phi$ given in (\ref{Sphi1}), once we specify our D3 brane configuration%
\footnote{As is clear from Fig. \ref{fig:phimu2mu3}, this contribution is present only when the
instanton is of stringy nature, {\it i.e.} $k_2=1$ or $k_3=1$, since the D$(-1)$ boundary must be of a different type with respect to the D3 boundaries. This type of contributions will play a crucial r\^ole in Section \ref{secn:strinst}, but
we discuss it here to illustrate in a simple example how the couplings with the holomorphic matter superfields
can be obtained.}.
Another possible diagram,
represented in Fig. \ref{fig:phimu2mu3}\emph{b)}, gives rise to the following coupling:
\begin{equation}
-\frac{\ii}{\sqrt2}\,\theta^\alpha\bar\mu^2\,\psi_\alpha(x)\mu^3~.
\label{psimumu}
\end{equation}
As discussed in \cite{Green:2000ke,Billo:2002hm,Billo:2006jm}, diagrams like those in Fig.
\ref{fig:phimu2mu3}\emph{a)} and \emph{b)} are related to each other by the action
of the two supersymmetries of the D-instanton broken by the D3 branes.
A further application of these supersymmetries leads to
\begin{equation}
 \frac{\ii}{2}\,\theta^2\,\bar\mu^2\,F(x)\,\mu^3~
\label{Fmumu}
\end{equation}
where $F$ is the auxiliary field of the matter multiplet. Also this coupling arises from a mixed disk diagram with two $\theta$-insertions on the D(--1) boundary and one insertion of $F$ on the
D3 boundary \footnote{For details on the calculations of disk amplitudes involving auxiliary fields
in this orbifold model see for example \cite{Billo:2005jw}.}. 
Adding the contributions (\ref{phimumu}), (\ref{psimumu}) and (\ref{Fmumu}), 
we reconstruct the combination
\begin{equation}
 \phi(x) + \sqrt2 \theta^\alpha\psi_\alpha(x) + \theta^2\,F(x)
\label{phipsi}
\end{equation}
which is the component expansion of the matter chiral 
superfield $\Phi(x,\theta)$ of our SQCD model.
Proceeding systematically in this way, one can show that the same pattern appears everywhere,
so that we can simply promote the vacuum expectation value $\phi$ to the complete superfield $\Phi(x,\theta)$, {\it i.e.} 
perform in the action (\ref{Sphi1}) the following replacement:
\begin{equation}
\phi~\to~\Phi(x,\theta)~
\label{promo0}
\end{equation}
in order to obtain all instanton couplings with the holomorphic scalar and its superpartners.
\begin{figure}[htb]
 \begin{center}
 \begin{picture}(0,0)%
\includegraphics{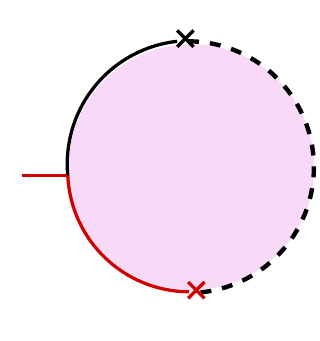}%
\end{picture}%
\setlength{\unitlength}{1381sp}%
\begingroup\makeatletter\ifx\SetFigFontNFSS\undefined%
\gdef\SetFigFontNFSS#1#2#3#4#5{%
  \reset@font\fontsize{#1}{#2pt}%
  \fontfamily{#3}\fontseries{#4}\fontshape{#5}%
  \selectfont}%
\fi\endgroup%
\begin{picture}(4344,4626)(496,-3859)
\put(2851,464){\makebox(0,0)[lb]{\smash{{\SetFigFontNFSS{8}{9.6}{\familydefault}{\mddefault}{\updefault}$\bar\mu$}}}}
\put(511,-1426){\makebox(0,0)[lb]{\smash{{\SetFigFontNFSS{8}{9.6}{\familydefault}{\mddefault}{\updefault}$\bar \phi$}}}}
\put(2926,-3736){\makebox(0,0)[lb]{\smash{{\SetFigFontNFSS{8}{9.6}{\familydefault}{\mddefault}{\updefault}$\mu^1$}}}}
\end{picture}%
 \end{center}
\caption{An example of a disk interaction between the (anti-holomorphic) scalar $\bar\phi(x)$ and the instanton moduli leading to the coupling (\ref{barphimumu}).}
\label{fig:psimumu1}
\end{figure}
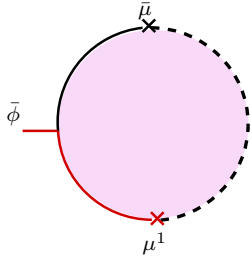 

Let us now turn to the anti-holomorphic variables. An example of a mixed disk
amplitude involving the scalar $\bar\phi(x)$ is represented in 
Fig. \ref{fig:psimumu1}.
It accounts for the coupling
\begin{equation}
-\frac{\ii}{2}\,\bar\mu\,\bar\phi(x)\,\mu^1~,
\label{barphimumu}
\end{equation}
which is the obvious generalization of the first term in the second line of
(\ref{Sphi1}) when the anti-holomorphic vacuum expectation value $\bar\phi$ is promoted
to a dynamical field.
The same pattern occurs in all
terms involving the anti-holomorphic vacuum expectation values $\bar\phi$, so that 
we can promote the latter with the replacement
\begin{equation}
\bar\phi~\to~\bar\phi(x) =\bar\Phi(x,\bar\theta)\Big|_{\bar\theta=0}~.
\label{promo00}
\end{equation}
Notice that no $\bar\theta$ dependence arises, due to the half-BPS nature of the D3/D$(-1)$ system.

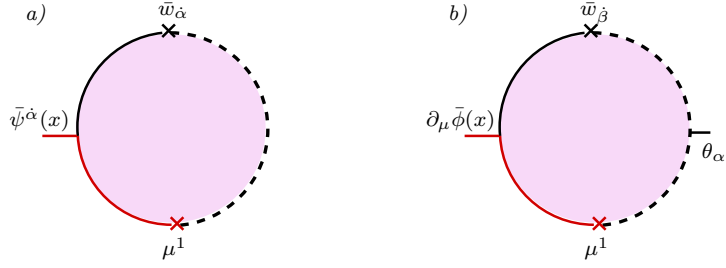
\begin{figure}[hbt]
 \begin{center}
  \begin{picture}(0,0)%
\includegraphics{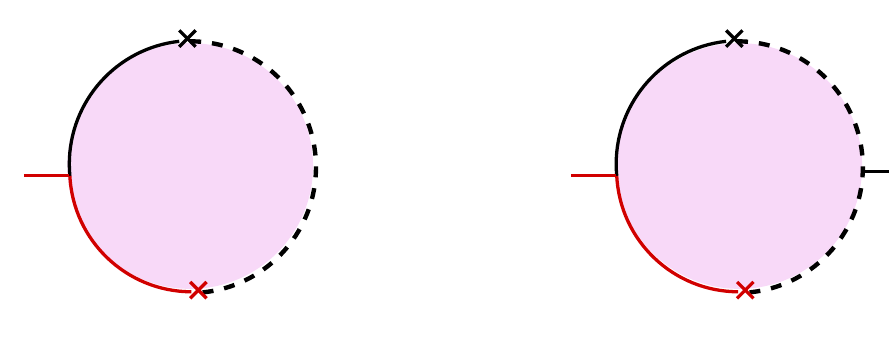}%
\end{picture}%
\setlength{\unitlength}{1381sp}%
\begingroup\makeatletter\ifx\SetFigFontNFSS\undefined%
\gdef\SetFigFontNFSS#1#2#3#4#5{%
  \reset@font\fontsize{#1}{#2pt}%
  \fontfamily{#3}\fontseries{#4}\fontshape{#5}%
  \selectfont}%
\fi\endgroup%
\begin{picture}(12228,4653)(466,-3859)
\put(10351,564){\makebox(0,0)[lb]{\smash{{\SetFigFontNFSS{8}{9.6}{\familydefault}{\mddefault}{\updefault}$\wdb$}}}}
\put(7611,-1426){\makebox(0,0)[lb]{\smash{{\SetFigFontNFSS{8}{9.6}{\familydefault}{\mddefault}{\updefault}$\partial_\mu\bar\phi(x)$}}}}
\put(10426,-3736){\makebox(0,0)[lb]{\smash{{\SetFigFontNFSS{8}{9.6}{\familydefault}{\mddefault}{\updefault}$\mu^1$}}}}
\put(481,479){\makebox(0,0)[lb]{\smash{{\SetFigFontNFSS{8}{9.6}{\familydefault}{\mddefault}{\updefault}\emph{a)}}}}}
\put(7996,464){\makebox(0,0)[lb]{\smash{{\SetFigFontNFSS{8}{9.6}{\familydefault}{\mddefault}{\updefault}\emph{b)}}}}}
\put(12511,-1996){\makebox(0,0)[lb]{\smash{{\SetFigFontNFSS{8}{9.6}{\familydefault}{\mddefault}{\updefault}$\ta$}}}}
\put(2851,564){\makebox(0,0)[lb]{\smash{{\SetFigFontNFSS{8}{9.6}{\familydefault}{\mddefault}{\updefault}$\wda$}}}}
\put(211,-1426){\makebox(0,0)[lb]{\smash{{\SetFigFontNFSS{8}{9.6}{\familydefault}{\mddefault}{\updefault}$\bar\psi^{\dot\alpha}(x)$}}}}
\put(2926,-3736){\makebox(0,0)[lb]{\smash{{\SetFigFontNFSS{8}{9.6}{\familydefault}{\mddefault}{\updefault}$\mu^1$}}}}
\end{picture}%
 \end{center}
\caption{Examples of disk diagrams responsible for the coupling of the superfield $D_{\dot\alpha}\bar\Phi(x,\bar\theta)\Big|_{\bar\theta=0}$ to the instanton moduli.
In \emph{b)} the vertex for the scalar $\bar\phi$ 
is in the 0-th superghost picture, which leads to a derivative coupling.}
\label{fig:psiwmu1}
\end{figure}

When one considers dynamical gauge fields, there
are new types of mixed disk amplitudes 
that correspond to couplings which do not depend on the vacuum expectation values of the scalars; 
since they are not present in the action (\ref{Sphi1}), they cannot be obtained with the
replacements (\ref{promo0}) and (\ref{promo00}).
These new types of interactions typically 
involve the D3/D3 anti-chiral fermions $\bar\psi_{\dot\alpha}(x)$ and
correspond to the following couplings:
\begin{equation}
\ii\,\bar w_{\dot \alpha}\bar\psi^{\,\dot\alpha}\!(x)\,\mu^1
-\ii\,\bar\mu^1\,\bar\psi_{\dot\alpha}(x)\,w^{\dot\alpha}~.
\label{wpsimu}
\end{equation}
Fig. \ref{fig:psiwmu1}\emph{a)} represents the disk diagram corresponding to the first term
of (\ref{wpsimu}). 
Furthermore, using the D3 supersymmetries that are broken by
the D$(-1)$ branes, we can produce the following terms:
\begin{equation}
-\theta_{\alpha}\bar w_{\dot \beta}(\bar\sigma^\mu)^{\alpha\dot\beta}
\partial_\mu\bar\phi(x)\,\mu^1
+\theta^{\alpha}\bar\mu^1\,(\bar\sigma^\mu)_{\alpha\dot\beta}
\partial_\mu\bar\phi(x)\,w^{\dot\beta}~.
\label{wpsimu2}
\end{equation}
The diagram responsible for the first term in (\ref{wpsimu2}) is represented in Fig. \ref{fig:psiwmu1}\emph{b)}.
The couplings (\ref{wpsimu2}) can be obtained from (\ref{wpsimu}) 
by means of the replacement
\begin{equation}
\bar\psi_{\dot\alpha}(x) ~\to~ \bar D_{\dot\alpha}\bar\Phi(x,\bar\theta)\Big|_{\bar\theta=0}~,
\label{promo2}
\end{equation}
where $\bar D_{\dot\alpha}$ is the standard spinor covariant derivative .  Note that
$\bar D_{\dot\alpha}\bar\Phi(x,\bar\theta)\Big|_{\bar\theta=0}$ is a chiral superfield \footnote{
Couplings
involving instanton moduli and chiral superfields of the form 
$\bar D_{\dot\alpha}\bar \Phi\big|_{\bar\theta=0}$ have been recently considered in Refs.
\cite{Matsuo:2008nu,GarciaEtxebarria:2008pi,Uranga:2008nh}.
}.

We are now  in the position of writing the action for the D3/D$(-1)$ system in 
presence of dynamical bi-fundamental matter fields, including string corrections.
In the case of a single gauge instanton configuration for the $\mathrm{SU}(N_0)$ factor, which corresponds to take $k_0=1$, this action is given by%
\footnote{Remember that in the one-instanton case for $\mathcal N=1$ models there are no
$\chi$-moduli.} 
\begin{equation}
\begin{aligned}
S_{\mathrm{D3/D(-1)}}(\Phi,\bar\Phi) = &~\frac{2\pi^3{\alpha'}^2}{g_s}\,D_cD^c+\ii\,D_{c}\big(\bar{w}_{\dot\alpha}
(\tau^c)^{\dot\alpha}_{~\dot\beta} w^{\dot\beta}\big)
+ \ii\,\lambda_{\dot\alpha}
\big( \bar{\mu}\,w^{\dot{\alpha}}
+\bar{w}^{\dot{\alpha}}\,\mu\big)
\\
&+\Big[\,\frac{1}{2}\,\bar{w}_{\dot\alpha}\big(\Phi\,\bar\Phi+
\bar\Phi\,\Phi\big) w^{\dot\alpha}+\frac{\ii}{2}\,\bar\mu^1\,
\bar\Phi\,\mu - \frac{\ii}{2}\,\bar\mu\, \bar\Phi\,\mu^1 \\
& ~~~~+\ii\,\bar w_{\dot \alpha}\big(\bar D^{\dot\alpha}\bar\Phi\big)\mu^1
-\ii\,\bar\mu^1\big(\bar D_{\dot\alpha}\bar\Phi\big) w^{\dot\alpha}
\Big]_{\bar\theta=0}~.
\end{aligned}
\label{sk2} 
\end{equation}
In the first line above the quadratic term in the auxiliary fields 
$D_c$ with an $\alpha'$-dependent coefficient comes from the gauge action $S_G$ in (\ref{Sd1d3})
written for a one-instanton configuration.
The second line in (\ref{sk2}) is the result of the replacements 
(\ref{promo0}) and (\ref{promo00}) in $S_\phi$, whereas
the third line arises from (\ref{wpsimu}) upon use of (\ref{promo2}).

In the following we will discuss the non-perturbative effective terms that are
induced on the D3 brane world volume by this instanton configuration.

\subsection{Field theory results: non-perturbative F-terms}
\label{subsecn:fieldtheory}

In the field theory limit $\alpha'\to 0$,
the instanton action (\ref{sk2}) simplifies to
\begin{equation}
\begin{aligned}
S^{(0)}_{\mathrm{D3/D(-1)}}(\Phi,\bar\Phi) & =
~\ii\,D_{c}\big(\bar{w}_{\dot\alpha}
(\tau^c)^{\dot\alpha}_{~\dot\beta} w^{\dot\beta}\big)
+\ii \,\lambda_{\dot\alpha} \big( \bar{\mu}\,w^{\dot{\alpha}}
+\bar{w}^{\dot{\alpha}}\,\mu\big)
\\
& +\Big[\,\frac{1}{2}\,\bar{w}_{\dot\alpha}\big(\Phi\,\bar\Phi+
\bar\Phi\,\Phi\big) w^{\dot\alpha}+\frac{\ii}{2}\,\bar\mu^1\,
\bar\Phi\,\mu - \frac{\ii}{2}\,\bar\mu\, \bar\Phi\,\mu^1 \\
&~~~~ +\ii\,\bar w_{\dot \alpha}\big(\bar D^{\dot\alpha}\bar\Phi\big)\mu^1
-\ii\,\bar\mu^1\big(\bar D_{\dot\alpha}\bar\Phi\big) w^{\dot\alpha}\Big]_{\bar\theta=0}~.
\end{aligned}
\label{sk1} 
\end{equation}
Note that in this action $D_c$ and $\lambda_{\dot\alpha}$ appear only linearly and act as Lagrange multipliers for the bosonic and fermionic ADHM constraints, and that,
as in (\ref{sk2}),
the dependence on the superspace coordinates 
$x^\mu$ and $\theta^\alpha$ is only through the matter superfields.

Integrating over all instanton moduli we obtain the following non-perturbative F-terms:
\begin{equation}
S_{\mathrm{n.p.}}= \int d^4x\,d^2\theta\,\,{W}_{\mathrm{n.p.}}
~,\quad
 {W}_{\mathrm{n.p.}}= \Lambda^{\beta_0} \int d\,{\widehat{\mathfrak M}}
\,~\ee^{-S^{(0)}_{\mathrm{D3/D(-1)}}(\Phi,\bar\Phi)}~,
\label{weff} 
\end{equation}
where
$\Lambda$ is the dynamically generated scale of the effective $\mathrm{SU}(N_0)$ SQCD theory we are considering, namely
\begin{equation}
 \Lambda^{\beta_0}=M_s^{\beta_0}\,\ee^{2\pi\ii\tau_0}
\quad\mbox{with}\quad \beta_0=3N_0-N_1~.
\end{equation}
Despite the notation we have adopted, one should not immediately
conclude that ${W}_{\mathrm{n.p.}}$ defined in (\ref{weff})
be a superpotential since, as we will see momentarily, gauge instantons
can induce also other types of non-perturbative F-terms.

In view of the explicit form of the field dependent moduli action (\ref{sk1}),
we can make the following general Ansatz:
\begin{equation}
 {W}_{\mathrm{n.p.}} = {\mathcal C}\, \Lambda^{\beta_0}\,
{\bar\Phi}^{n}\,\Phi^{m}\,\big({\bar D_{\dot\alpha}
\bar\Phi\,\bar D^{\dot\alpha}\bar\Phi}
\big)^{p}\Big|_{\bar\theta=0}~,
\label{ansatz1}
\end{equation}
where $p$ is restricted to positive values to avoid the appearance of fermionic fields in the 
denominator. We now proceed as in Section \ref{subsec:zeromode} and require
${W}_{\mathrm{n.p.}}$ to 
be a quantity of scaling dimension 3 and that its $\mathrm U(1)^3$ charges match 
those of the centered measure, given in (\ref{qqq}) with $k_1=0$. Taking into account 
that $q[\bar D\bar\Phi]=+1/2$ and 
$q^\prime[\bar D\bar\Phi]=q^{\prime\prime}[\bar D\bar\Phi]=-1$,
after some simple algebra we find that the parameters in (\ref{ansatz1}) are given by
\begin{equation}
p=-n= 1-N_0+N_1~,\quad m=1-N_0-N_1~.
\label{pmn}
\end{equation}
The instanton induced effective interactions have thus the form
\begin{equation}
{W}_{\mathrm{n.p.}} = {\mathcal C}\, \Lambda^{\beta_0}\,
\frac{\big({\bar D_{\dot\alpha}\bar\Phi\,\bar D^{\dot{\alpha}}
\bar\Phi}\big)^{p}}{{\bar\Phi}^{p}\,\Phi^{\,p+2N_0-2}}\Bigg|_{\bar\theta=0}
\label{weff2}
\end{equation}
for $p=0,1,\ldots$.

For $p=0$ (and hence for $N_1=N_0-1$) the above result reduces to the
well-known ADS superpotential for $\mathrm{SU}(N_c)$
SQCD with $N_f=N_c-1$ \cite{Affleck:1983mk}; indeed,
after using the D-flatness condition on the matter fields and
explicitly performing the integrations over all ADHM moduli in this case, 
one can prove that the overall coefficient $\mathcal C$ is non-vanishing and that
(\ref{weff2}) becomes
\begin{equation}
{W}_{\mathrm{n.p.}} =\mathcal C\,\Lambda^{2N_c+1}\,\frac{1}{\det M}
\label{wads}
\end{equation}
where $M$ is the meson superfield, in agreement with the ADS result.

For $p>0$ the above result (\ref{weff2}) reproduces 
the multi-fermion instanton induced interactions for SQCD with $N_f\geq N_c$ 
studied originally in \cite{Beasley:2004ys} in the case
$N_c=2$ and recently derived by integrating the ADHM moduli in \cite{Matsuo:2008nu} %
\footnote{See also Ref. \cite{GarciaEtxebarria:2008pi} for related considerations in the
case $N_c=1$.}. In particular, for $p=1$ and $N_c=N_f=2$, Eq. (\ref{weff2}) yields the form
\begin{equation}
{W}_{\mathrm{n.p.}}= {\mathcal C}\, \Lambda^{4}\,
\frac{{\bar D_{\dot\alpha}\bar\Phi\,\bar D^{\dot{\alpha}}
\bar\Phi}}{{\bar\Phi}\,\Phi^3}\Bigg|_{\bar\theta=0}~,
\label{BW}
\end{equation}
in accordance with the explicit result of the moduli integral 
\cite{Beasley:2004ys,Matsuo:2008nu}, which can be written as
\begin{equation}
{W}_{\mathrm{n.p.}}= \mathcal C\,\Lambda^{4}\,
\frac{\epsilon_{{f_1}f'_1}\,
\epsilon^{f_2f'_2}\,\bar D_{\dot\alpha}{\bar M}^{f_1}_{~f_2}\,
\bar D^{\dot{\alpha}}{\bar M}^{f'_1}_{~f'_2} +2\,\bar D_{\dot\alpha}\bar B
\bar D^{\dot\alpha}\bar{\tilde B}}{\big(\tr \bar M M+\bar B B+ \bar {\tilde B}\tilde B\big)^{3/2}}\Bigg|_{\bar\theta=0}~,
\label{BW2}
\end{equation}
in terms of the $\mathrm{SU}(2)$ meson and baryon fields 
(see Eq.s (\ref{meson}) and (\ref{baryon}) for $N_f=N_c=2$). 
For $p>1$ one obtains more general multi-fermion terms.
As proved in Ref. \cite{Beasley:2004ys}, these multi-fermion terms, despite
being non-holomorphic in the matter fields,
are annihilated by the anti-chiral supercharges $\overline Q_{\dot\alpha}$, and as such they are genuine F-terms even if they do not correspond to a superpotential.

\section{Non-perturbative flux-induced effective interactions}
\label{secn:fluxeffects}

\begin{figure}[htb]
 \begin{center}
  \begin{picture}(0,0)%
\includegraphics{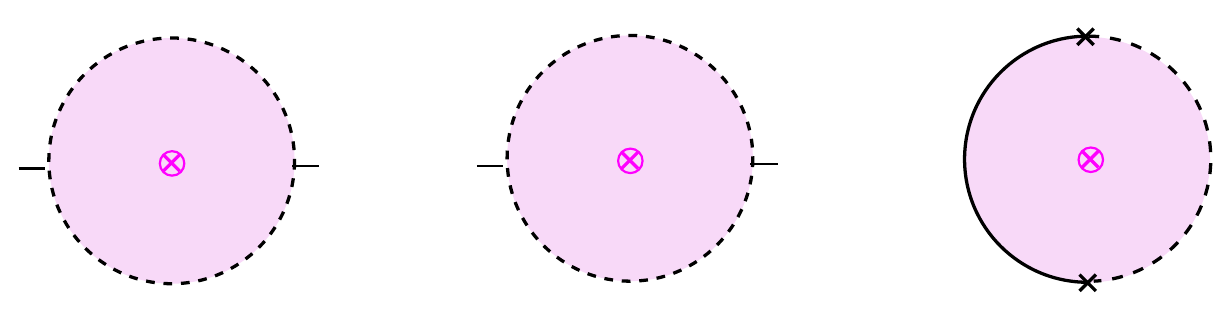}%
\end{picture}%
\setlength{\unitlength}{1381sp}%
\begingroup\makeatletter\ifx\SetFigFontNFSS\undefined%
\gdef\SetFigFontNFSS#1#2#3#4#5{%
  \reset@font\fontsize{#1}{#2pt}%
  \fontfamily{#3}\fontseries{#4}\fontshape{#5}%
  \selectfont}%
\fi\endgroup%
\begin{picture}(16636,4446)(796,-3754)
\put(11236,-1976){\makebox(0,0)[lb]{\smash{{\SetFigFontNFSS{8}{9.6}{\familydefault}{\mddefault}{\updefault}$\lda$}}}}
\put(7096,-1391){\makebox(0,0)[lb]{\smash{{\SetFigFontNFSS{8}{9.6}{\familydefault}{\mddefault}{\updefault}$\ldaup$}}}}
\put(15301,389){\makebox(0,0)[lb]{\smash{{\SetFigFontNFSS{8}{9.6}{\familydefault}{\mddefault}{\updefault}$\bar\mu$}}}}
\put(15421,-3631){\makebox(0,0)[lb]{\smash{{\SetFigFontNFSS{8}{9.6}{\familydefault}{\mddefault}{\updefault}$\mu$}}}}
\put(4951,-2011){\makebox(0,0)[lb]{\smash{{\SetFigFontNFSS{8}{9.6}{\familydefault}{\mddefault}{\updefault}$\taup$}}}}
\put(811,-1426){\makebox(0,0)[lb]{\smash{{\SetFigFontNFSS{8}{9.6}{\familydefault}{\mddefault}{\updefault}$\ta$}}}}
\put(2536,-1131){\makebox(0,0)[lb]{\smash{{\SetFigFontNFSS{8}{9.6}{\familydefault}{\mddefault}{\updefault}$G_{(3,0)}$}}}}
\put(9001,-1066){\makebox(0,0)[lb]{\smash{{\SetFigFontNFSS{8}{9.6}{\familydefault}{\mddefault}{\updefault}$G_{(0,3)}$}}}}
\put(15256,-1081){\makebox(0,0)[lb]{\smash{{\SetFigFontNFSS{8}{9.6}{\familydefault}{\mddefault}{\updefault}$G_{(3,0)}$}}}}
\end{picture}%
 \end{center}
\caption{Diagrams encoding the linear couplings of bulk fluxes of type $(0,3)$ and $(3,0)$ to the instanton moduli reported in Eq. (\ref{sflux1}).}
\label{fig:fluxed}
\end{figure}
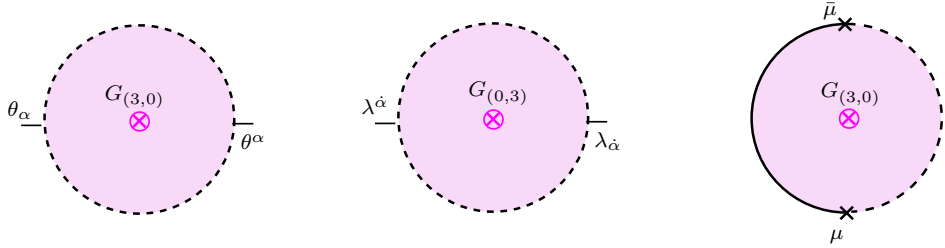

We now generalize the analysis of the previous sections and investigate the
non-perturbative effects produced in the gauge theory by adding R-R and NS-NS fluxes
in the internal space.
In particular we consider the 3-form flux
\begin{equation}
 G_3 = F - \tau \,H
\end{equation}
which is made out of the R-R 3-form $F$, the NS-NS 3-form $H$, and the axion-dilaton $\tau$. 
The general couplings of 
closed string fluxes to open string fermionic bilinears were derived in \cite{Billo:2008sp}
by evaluating mixed open/closed string amplitudes on disks with generalized mixed
boundary conditions like those represented in Fig. \ref{fig:fluxed}. 
We refer to the above reference for a detailed discussion; here we simply 
recall that this world-sheet approach is particularly suited to study the case at hand and discuss
how the $G_3$-fluxes couple to the instanton moduli%
\footnote{Notice that, differently from Ref. \cite{Billo:2008sp}, here the transverse space
to the D3-branes is non-compact. This implies that the normalization coefficients
to be used here are like those computed in Ref. \cite{Billo:2008sp}, but without the factors of
the compactification volume $\mathcal V$.}.
Indeed, computing the mixed open/closed string diagrams of Fig. \ref{fig:fluxed} 
in the $\mathbb Z_2 \times \mathbb Z_2$ orbifold, one finds that
the flux induced interactions on the instanton moduli space
are encoded in the action 
\begin{equation}
S^{\mathrm{flux}} = {2\pi\ii}\, \left[
\frac{2G_{(3,0)}}{\sqrt{g_s}}\,\theta^{\alpha} \theta_{\alpha}
-\frac{2G_{(0,3)}}{\sqrt{g_s}}\,\frac{\pi^2{\alpha'}^2}{2}\,\lambda_{\dot\alpha}\lambda^{\dot\alpha}
\right]
+{\ii}\sqrt{g_s}\,G_{(3,0)}\,{\bar\mu}\mu~,
\label{sflux1}
\end{equation}
where we have denoted by $G_{(3,0)}$ and $G_{(0,3)}$ the $(3,0)$ and $(0,3)$ components 
of $G_3$ in the natural complex structure of the transverse space. 
These components satisfy, respectively, an imaginary self-duality and anti-self-duality 
condition and are responsible for the soft supersymmetry breaking terms 
related to the gravitino and gaugino masses, see {\it e.g.} \cite{Camara:2003ku,Camara:2004jj}.
Note that in the first and last
terms of (\ref{sflux1}) the scaling dimension of (length)$^{-1}$ carried by the $G$-flux is 
compensated by the dimensions of $\theta$, $\mu$ and $\bar\mu$, while in the second term explicit $\alpha'$ factors are needed. This is perfectly consistent with the fact that, while the $G_{(3,0)}$ flux components have a natural field theory interpretation as gaugino masses, the 
$G_{(0,3)}$ components instead have no counterpart on the gauge field theory. Thus, from the open
string point of view their presence in (\ref{sflux1}) is a genuine string effect, 
as revealed also by the explicit factors of $\alpha'$.

The action (\ref{sflux1}) can be conveniently rewritten as
\begin{equation}
S^{\mathrm{flux}} = \frac{2\pi\ii}{g_s}\, \Big[\,G\,\theta^{\alpha} \theta_{\alpha}
-\bar G\,\frac{\pi^2{\alpha'}^2}{2}\,\lambda_{\dot\alpha}\lambda^{\dot\alpha}
\Big]
+\frac{\ii}{2}\,G\,{\bar\mu}\mu~,
\label{sflux2}
\end{equation}
where we have defined
\begin{equation}
 G = 2\sqrt{g_s} \,G_{(3,0)}~,\quad
\bar G = 2\sqrt{g_s} \,G_{(0,3)}~.
\label{GbarG}
\end{equation}
Eq. (\ref{sflux2}) is the form of the flux induced moduli action which we
will use in the following to study the non-perturbative interactions generated by
fractional D-instantons in the presence of bulk fluxes.
In particular we will consider terms at the linear order in $G$ or $\bar G$ where the methods of
\cite{Billo:2008sp} for the world-sheet derivation of the moduli action (\ref{sflux2})
are reliable.
We therefore have two possibilities depending on whether we keep $G$ or $\bar G$
different from zero, which we are going to analyze in turn.

\subsection{One-instanton effects with $G\neq 0$}
\label{subsecn:Gflux}
In this case we can set $\bar G=0$ and look for the non-perturbative interactions proportional
to $G$, assuming again that the fractional D-instanton is of type 0, {\it i.e.} that $k_0=1$,
as in Section \ref{secn:1inst}.
A class of such interactions is obtained by exploiting the $\frac{\ii}{2}G\bar\mu\mu$ term of the
flux action (\ref{sflux2}). At first order in $G$ this leads to
\begin{equation}
S_{\mathrm{n.p.}}(G) = 
\Lambda^{\beta_0}\int d^4x\,d^2\theta
\,d\,{\widehat{\mathfrak M}}
~\ee^{-S_{\mathrm{D3/D(-1)}}(\Phi,\bar\Phi)}\,\left(\frac{\ii}{2}\,G\bar\mu\,\mu\right)
\label{seffg}
\end{equation}
where $S_{\mathrm{D3/D(-1)}}(\Phi,\bar\Phi)$ is the instanton action (\ref{sk2}).
By taking the limit $\alpha'\to 0$ we
obtain non-perturbative flux-induced terms in the effective action of the form
\begin{equation}
S_{\mathrm{n.p.}}(G) = \int d^4x\,d^2\theta \,{W}_{\mathrm{n.p.}}(G)
~,~~~
{W}_{\mathrm{n.p.}}(G) = \Lambda^{\beta_0}
\int d\,{\widehat{\mathfrak M}}
\,~\ee^{-S^{(0)}_{\mathrm{D3/D(-1)}}(\Phi,\bar\Phi)}\,\left(\frac{\ii}{2}\,G\bar\mu\,\mu\right)
\label{weffg}
\end{equation}
where $S^{(0)}_{\mathrm{D3/D(-1)}}(\Phi,\bar\Phi)$ is the moduli action in the field theory
limit given in (\ref{sk1}).
After performing the integration over all centered moduli,
in the effective field theory we expect to find an interaction of the following 
schematic form
\begin{equation}
{W}_{\mathrm{n.p.}}(G)= {\mathcal C}\, G\,\Lambda^{\beta_0}\,
{\bar\Phi}^{n}\,\Phi^{m}\,\big({\bar D_{\dot\alpha}
\bar\Phi\,\bar D^{\dot\alpha}\bar\Phi}
\big)^{p}\Big|_{\bar\theta=0}
\label{weffg2}
\end{equation}
with $\beta_0+n+m+3p=2$ in order to have an operator of mass dimension 3 (remember that
$G$ has dimensions of a mass). As before, we restrict to positive values of $p$ in order to
avoid the appearance of fermionic fields in the denominator. Requiring that the three $\mathrm U(1)$ charges of ${W}_{\mathrm{n.p.}}(G)$ match those of 
the centered instanton measure for consistency with (\ref{weffg}), and using the
information that $q(G)=-3$ and $q^\prime(G)=q^{\prime\prime}(G)=0$, it is easy to find that
the parameters in (\ref{weffg2}) are given by
\begin{equation}
 p=-n-2=2-N_0+N_1\quad\mbox{and}\quad m=-N_0-N_1~.
\label{mnpG}
\end{equation}
The resulting multi-fermion interactions are non-supersymmetric as can
be easily seen by noticing that they are non-holomorphic for $p=0$.

The case $p=1$ ({\it i.e.} $N_1=N_0-1$) is particularly interesting, 
since it corresponds to $\mathrm{SU}(N_c)$ SQCD with $N_f=N_c-1$. 
We have already recalled that in this case the
gauge instanton induces the ADS superpotential; now we see that in the presence of a $G$-flux
which softly breaks supersymmetry by giving a mass to the gaugino, the gauge instanton produces
new types of low-energy effective interactions which are of the form
\begin{equation}
 {W}_{\mathrm{n.p.}}(G)= \left.{\mathcal C}\, G\,\Lambda^{2N_c+1}\,
\frac{{\bar D_{\dot\alpha}
\bar\Phi\,\bar D^{\dot\alpha}\bar\Phi}}{\bar\Phi^3\,\Phi^{2N_c-1}}\right|_{\bar\theta=0}~.
\label{weffG1}
\end{equation}
We stress that this is a formal expression which only indicates the powers of the various
fields that appear in the result; the precise structure of ${W}_{\mathrm{n.p.}}(G)$ 
should be given in terms of the appropriate variables of the low-energy effective theory
(the meson superfields in this case) and can be obtained by explicitly performing the integral
over the instanton moduli which also dictates how color and flavor indices must be saturated. 
This task is particularly easy to do for $\mathrm{SU}(2)$ SQCD with
one flavor, and some details can be found in Appendix \ref{app:intG}. There we show that for
$N_c=2$ and $N_f=1$ the flux induced non-perturbative term (\ref{weffG1}) can be written as
\begin{equation}
{W}_{\mathrm{n.p.}}(G)= \left. \mathcal C\,G\,\Lambda^5\,\frac{\bar D^2 \bar M}{(\bar M M)^{3/2}}
\right|_{\bar\theta=0}
\label{N2G}
\end{equation}
where $M$ is the meson superfield of the effective theory. We can regard this interaction
as a low-energy non-perturbative effect of the soft supersymmetry breaking realized by the
$G$-flux in the microscopic high-energy theory.
Finally, we observe that one can alternatively exploit the
$G\theta^2$ term of the  flux action (5.3) to produce non-supersymmetric
interactions of the same type as the ones described here.

\subsection{One-instanton effects with $\bar G \neq 0$}
\label{subsecn:barGflux}
The contribution to the effective action linear in $\bar G$ in presence of a single fractional D-instanton of type 0 is given, in analogy to Eq. (\ref{seffg}), by
\begin{equation}
S_{\mathrm{n.p.}}(\bar G) = \Lambda^{\beta_0}\,\int d^4x\,d^2\theta
\,d^2\lambda\,\,d\,{\widehat{\mathfrak M}'}
~\ee^{-S_{\mathrm{D3/D(-1)}}(\Phi,\bar\Phi)}\,
\left(-\frac{2\pi\ii}{g_s}\,\frac{(\pi\alpha')^2}{2}\,\bar G \lambda_{\dot\alpha}\lambda^{\dot\alpha}\right)~.
\label{seffgbar}
\end{equation}
Here we have denoted by ${\widehat{\mathfrak M}'}$ all centered moduli but $\lambda$. 
Performing the Grassmannian integration over $d^2\lambda$, we can write
\begin{equation}
S_{\mathrm{n.p.}}(\bar G) = \int d^4x\,d^2\theta
\,\,W_{\mathrm{n.p.}}(\bar G)~,
\label{seffgbar2}
\end{equation}
where
\begin{equation}
W_{\mathrm{n.p.}}(\bar G)=(\pi\alpha')^2 \,\frac{2\pi\ii}{g_s}\,\Lambda^{\beta_0}\,\bar G\, 
\int d\,{\widehat{\mathfrak M}'} \,~\ee^{-\left.S'_{\mathrm{D3/D(-1)}}(\Phi,\bar\Phi)\right.
}
\label{weffgbar}
\end{equation}
with $S'_{\mathrm{D3/D(-1)}}(\Phi,\bar\Phi)$ being the action (\ref{sk2}) without the fermionic
ADHM constraint term since the Grassmannian integration over $\lambda$ has killed it. {From} (\ref{weffgbar})
we therefore expect to find a result of the schematic form
\begin{equation}
{W}_{\mathrm{n.p.}}(\bar G) = {\mathcal C}\,\alpha'^2\, \bar G\,\Lambda^{\beta_0}\,
{\bar\Phi}^{n}\,\Phi^{m}\,\big({\bar D_{\dot\alpha}
\bar\Phi\,\bar D^{\dot\alpha}\bar\Phi}
\big)^{p}\Big|_{\bar\theta=0} + \cdots~,
\label{weffgbar2}
\end{equation}
where the dots stand for possible higher order string corrections.
Requiring the equality of dimensions and $\mathrm{U}(1)^3$ charges between the definition
(\ref{weffgbar}) and the expression (\ref{weffgbar2}), one finds
\begin{equation}
 \label{parbar}
m= 3 -N_0 - N_1\quad,\quad n = 3 + N_0 - N_1\quad,\quad p = N_1 - N_0~.
\end{equation}

Let us focus on the simple case $p=0$ ({\it i.e.} $N_0=N_1$), which corresponds to a SQCD with $N_f=N_c$ flavors. 
In this case, in absence of fluxes, one gets only multi-fermion terms of Beasley-Witten type, 
like the ones displayed in Eq.s (\ref{BW}) and (\ref{BW2}). In the presence of a $\bar G$-flux
we have also a non-holomorphic contribution of the form
\begin{equation}
 \label{cp0}
{W}_{\mathrm{n.p.}} = \mathcal{C} \,\alpha'^2\, \bar G\,\Lambda^{2 N_c}\,
{\bar\Phi}^{3}\,\Phi^{3 - 2 N_c}\Big|_{\bar\theta=0}
\end{equation}
which can be explicitly computed by performing the integration
over  $d\,{\widehat{\mathfrak M}'}$, as shown in
Appendix \ref{app:intGbar}. Notice that again these terms are non-holomorphic and therefore manifestly
non-supersymmetric. For $N_c=2$, the result is
\begin{equation}
 \label{GbarN2}
{W}_{\mathrm{n.p.}} = \mathcal{C} \,\alpha'^2\, \bar G\,\Lambda^{4}\,
\left.\frac{\det \bar M}{\big(\tr \bar M M+\bar B B+ \bar {\tilde B}\tilde B\big)^{1/2}} \right|_{\bar\theta=0}~,
\end{equation}
where $M$ is the meson superfield and $B$ and $\widetilde B$ are the baryon superfields.

\section{Stringy instanton effects in presence of fluxes}
\label{secn:strinst}
D-instantons of type $2$ and $3$ are of different type with respect to
the D3 branes where the SQCD-like $\mathrm{SU}(N_0)\times \mathrm{SU}(N_1)$ theory
is defined, and lead thus to ``stringy'' or ``exotic'' non-perturbative effects.
In this case we have only fermionic mixed moduli $\mu^2$, $\bar\mu^2$ and
$\mu^3$, $\bar\mu^3$, while there are no $w_{\dot\alpha}$ and $\bar w_{\dot\alpha}$'s
from the NS sectors. 

Let us now derive the general form of the centered partition function
in dependence of the vacuum expectation value of the scalar $\phi$, in analogy to
what we did for the gauge instantons in Section \ref{sec:effint}.
The moduli action (\ref{cometipare}) drastically
simplifies. In particular, it is holomorphic in $\phi$
and does not contain any $\lambda$ dependence.
Therefore, unless one introduces an orientifold projection \cite{Argurio:2007qk,Argurio:2007vqa,Bianchi:2007wy}
or invokes other mechanisms \cite{Blumenhagen:2007bn,Petersson:2007sc,GarciaEtxebarria:2008pi},
the only way to get a non-zero result is to include
the flux-induced $\bar G\lambda\lambda$ term of Eq. (\ref{sflux1}) and
use it to perform the $\lambda$ integration. At the linear level in the fluxes,
the other flux interactions in (\ref{sflux2}) become then irrelevant.
Neglecting as usual numerical prefactors, we can write the centered partition function for
a stringy instanton configuration with instanton numbers $k_2$ and $k_3$ as
\begin{equation}
 \label{Wsi}
W_{\mathrm{n.p.}}(\bar G) = \alpha'^2\,\bar G\,M_s^{k_2\beta_2 +
k_3\beta_3}\,\ee^{2\pi\ii(k_2\tau_2 + k_3\tau_3)}\int d\,{\widehat{\mathfrak
M}'}\, \ee^{-S_{\mathrm{D3/D(-1)}}}~,
\end{equation}
leading to the following general Ansatz 
\begin{equation}
 \label{ansWsi}
W_{\mathrm{n.p.}}(\bar G)= \mathcal C\,\bar G\, M_s^{k_2\beta_2 +
k_3\beta_3 + n}\,\ee^{2\pi\ii(k_2\tau_2 + k_3\tau_3)}\, \phi^m~. 
\end{equation}
We have not fixed a priori the power of $M_s\sim 1/\sqrt{\alpha'}$ since
in this case the moduli integration can produce extra factors of $\alpha'$ with
respect to those appearing in Eq. (\ref{Wsi}), because of the $S_G$ part of the
moduli action (\ref{cometipare}) which appears with an explicit $(\alpha')^2$ in
front%
\footnote{In particular, the ``center of mass'' part of the $D_c$'s appears
only through the quadratic term $\sim\alpha'^2D_c D^c$ and the gaussian
integration over it produces negative powers of $\sqrt{\alpha'}$. This is different
with respect to the gauge instanton cases considered in Sections \ref{secn:1inst} and 
\ref{secn:fluxeffects}, where the $D_c$'s couple also to the bosonic moduli $w$ and
$\bar w$, leading to a completely different type of integral.}.

The equality between the $q,q'$ and $q''$ charges following from this ansatz
and those implied by the definition (\ref{Wsi}) plus the request
that the mass dimension of $W_{\mathrm{n.p.}}$ be equal to 3 impose that
\begin{equation}
\label{sicond}
 n= 2\quad,\quad m = 2(k_2 N_0 + k_3 N_1)~,
\end{equation}
and
\begin{equation}
\label{k2k3N0N1}
 (k_2 - k_3)(N_0 - N_1)  = 0~;
\end{equation}
To derive these equations we have used the fact that for our brane configuration
$\beta_2 = \beta_3 = -(N_0 + N_1)$.
The condition (\ref{k2k3N0N1}) admits the following solutions:
\begin{equation}
 \label{sisol}
\begin{aligned}
 k_2 &= k_3\quad\mbox{with}\quad N_0~\mbox{and}~N_1~\mbox{arbitrary}~,\\
 N_0 &= N_1\quad\mbox{with}\quad k_2~\mbox{and}~k_3~\mbox{arbitrary}~. 
\end{aligned}
\end{equation}
The centered partition function in stringy instanton sectors can thus be written
in the form
\begin{equation}
 \label{Wsires}
W_{\mathrm{n.p.}} (\bar G)= \mathcal C\,
\bar G\, M_s^{k_2\beta_2 + k_3\beta_3 + 2}\,\ee^{2\pi\ii(k_2\tau_2 + k_3\tau_3)}\,
\phi^{-(k_2\beta_2 + k_3\beta_3)}~.
\end{equation}

Let us now concentrate on the set-up containing a single stringy instanton
described in Fig. \ref{fk0k2}\emph{b)}, namely let us set $k_2=1$, $k_3=0$.
In this case it is easy to promote the vacuum expectation value $\phi$ 
to the full superfield $\Phi(x,\theta)$ through
diagrams such as those of Fig. \ref{fig:phimu2mu3}.
and the moduli integration can be explicitly done. 
As remarked above, 
the only way to saturate the Grassmannian integration over
$d\lambda_{\dot\alpha}$ is via their $\bar G$ interaction
and the non-perturbative contribution to the effective action of this 
``stringy'' instanton sector is
\begin{equation}
 \label{Seffsi}
S_{\mathrm{n.p.}} = 
\int d^4x\,d^2\theta\, W_{\mathrm{n.p.}}(\bar G)
\end{equation}
where the superpotential is given by
\begin{equation}
{W_{\mathrm{n.p.}}}= \mathcal C\,{\alpha'}^2\,
M_s^{-(N_0 + N_1)}\,\ee^{2\pi\ii\tau_2}\,\bar G\,
\int d\,{\widehat{\mathfrak M}'}
\,\ee^{-S_{\mathrm{D3/D(-1)}}(\Phi)}~.
\label{Weffsi}
\end{equation}
Notice that the dimensional prefactor does not combine
with the exponential of the classical action  to form the
dynamically generated scale of the gauge theory, since
$\tau_2$ is the complexified coupling of D3-branes of type 2, which are not the ones that
support the gauge theory we are considering.

The moduli $\widehat{\mathfrak M}'$ appearing in (\ref{Weffsi}) are
simply $\{D_c,\mu^2\,{\bar\mu}^2,\mu^3\,{\bar\mu}^3\}$, with $\mu^2$ and $\mu^3$
transforming in the fundamental representations of
$\mathrm{U}(N_0)$ and $\mathrm{U}(N_1)$ respectively. Thus, the moduli action to be used in (\ref{Weffsi}) simply 
reduces to
\begin{equation}
 \label{seffex}
S_{\mathrm{D3/D(-1)}}(\Phi) =
\frac{2\pi^3{\alpha'}^2}{g_s}\,D_cD^c -\frac{\ii}{2} 
\left({\bar\mu}^3\Phi\mu^2 -{\bar\mu}^2\Phi\mu^3 \right)~. 
\end{equation}
Hence, the integral in (\ref{Weffsi}) explicitly reads
\begin{equation}
 \label{modintsi}
\int d^3D\,d^{N_0}\mu^2\, d^{N_0}{\bar\mu}^2\, d^{N_1}\mu^3\, d^{N_1}{\bar\mu}^3
\ee^{-\frac{2\pi^3{\alpha'}^2}{g_s}\,D_cD^c +\frac{\ii}{2} 
\left({\bar\mu}^3\Phi\mu^2 -{\bar\mu}^2\Phi\mu^3 \right) }~.
\end{equation}
The integration over the $\mu$'s clearly vanishes unless $N_0=N_1$, in which
case we get, after performing also the gaussian integration over the $D$'s,
\begin{equation}
 \label{risintsi}
{\alpha'}^{-3} \det Q\, \det \tilde Q =
{\alpha'}^{-3} \det M~.
\end{equation}
Here we have used the form (\ref{Phi}) of $\Phi$ and in the last step we have introduced
the meson field $M=\tilde Q Q$. We have also disregarded all numerical constants and kept track
only of the powers of $\alpha'\propto M_s^{-2}$, since in all of
our treatment we have specified completely only the dimensional part of the prefactors 
in the moduli measure.

Inserting (\ref{risintsi}) into (\ref{Weffsi}) we find therefore that a single
stringy instanton in presence of an imaginary self-dual three-form flux produces for $N_0=N_1$ 
({\it i.e.} for a SQCD with $N_f=N_c$ flavors) a
\emph{holomorphic} superpotential
\begin{equation}
 \label{risweffsi}
W_{\mathrm{n.p.}}= \mathcal{C}\, M_s^{2 - 2
N_c}\ee^{2\pi\ii\tau_2}\, \bar G\, \det M~.
\end{equation}
Interestingly, the interactions generated by stringy instantons
are still holomorphic and therefore supersymmetric even in the
presence of the supersymmetry breaking flux $\bar{G}$. it would be interesting to investigate
the implications of such non-perturbative terms for the low-energy effective action.
\vskip 1cm
\noindent {\large {\bf Acknowledgments}}
\vskip 0.2cm
\noindent We thank B. Acharya, C. Angelantonj, C. Bachas, K.S. Narain, I. Pesando, R. Russo and T. Weigand for
useful discussions.
We furthermore thank the referee for questions and clarifying comments.
This work is partially supported by the European Commission FP6 Programme under contract MRTN-CT-2004-005104 ``{Constituents, Fundamental Forces and Symmetries of the Universe}'', in which
A.L. is associated to University of Torino, MRTN-CT-2004-512194 ``{Superstring Theory}'' and  MRTN-CT-2004-503369 ``{The Quest for Unification: Theory Confronts Experiment}'', by the Italian MIUR-PRIN contract  20075ATT78 and by the NATO grant PST.CLG.978785.
L.F. would like to thank C. Bachas for kind support and 
LPTENS for wonderful hospitality. 
\vskip 1cm
\appendix

\section{Appendix}
\subsection{Gauge coupling and instanton action}
\label{apDBI}
Let us consider a D$p$-brane wrapping a $(p-3)$-cycle $\mathcal C_A$ and denote by
$\tau_A$ the complexified gauge coupling of the resulting four-dimensional super Yang-Mills theory:
\begin{equation}
\tau_A=\frac{\theta_A}{2\pi}+\ii\,\frac{4\pi}{g_A^2}
\end{equation}
A gauge instanton in this theory can be described in terms of a Euclidean $(p-4)$-brane 
wrapping the same $(p-3)$-cycle $\mathcal C_A$. The instanton induces non-perturbative interactions weighted by $\ee^{-k_A S_A^{\mathrm E(p-4)}}$ with $k_A$ being
the number of instantonic branes and $S_{A}$ the action for a single instanton. 
Aim of this appendix is to show the relation
\begin{equation}
S_{A}^{\mathrm E(p-4)}=-2\pi \ii \tau_A~. 
\label{sinst}
\end{equation}
which justifies the form of $W_{\mathrm{n.p.}}$ in (\ref{wfluxnp}).

Eq. (\ref{sinst}) follows from a comparison of the world-volume action of the Euclidean
E$(p-4)$-brane with that of the wrapped D$p$-brane \cite{Billo:2007sw}.
In Euclidean signature, the latter is%
\footnote{Here we assume $\big[F_{\mu\nu},F_{\sigma \rho}\big]=0$ and take
$F=F_i T^j$ with   ${\mathrm{Tr}}\big(T^i T^j\big)=\frac{1}{2} \delta^{ij}$ and $i,j$
running in the adjoint of the gauge.}
\begin{equation}
S_A^{\mathrm{D}p} = \mu_{p} \,{\mathrm{Tr}}\left[\int_{\mathbb R^4\times \mathcal C_A}
\!\!\ee^{-\varphi}\,\sqrt{\,\det \,\big( g+2\pi \alpha' F\big)} ~-~
\ii\int_{\mathbb R^4\times \mathcal C_A}\,\sum_{n}\, C_{2n} \,\ee^{2\pi \alpha' F}\right]~,
\label{cal}
\end{equation}
where $\mu_p=(2\pi)^{-p} (\alpha')^{-(p+1)/2}$ is the D$p$-brane tension, $\varphi$ the
dilaton, $g$ the string frame metric and $C_{2n}$ the R-R $2n$-form potentials.
Expanding (\ref{cal}) to quadratic order in $F$ and comparing with the standard
form of the Yang-Mills action in Euclidean signature, we find
that the complexified four-dimensional gauge coupling is 
\begin{equation}
\tau_A=  2\pi (2\pi \alpha')^2 \, \mu_{p}\,\int_{\mathcal C_A}
\left[ C_{p-3}+\ii \, \ee^{-\varphi}\,
\sqrt{\,\det \,g }  \right]~. 
\label{taua}
\end{equation}
On the other hand the action for a Euclidean $(p-4)$-brane
wrapping $\mathcal C_A$ is given by
\begin{equation}
S_A^{\mathrm{E}(p-4)} =\mu_{p-4} \left[\int_{\mathcal C_A} \ee^{-\varphi}\,
\sqrt{\,\det \,g } ~-~\ii \int_{\mathcal C_A}\, C_{p-3}\right] =-2\pi\ii\, \tau_A
   \label{cal2}
  \end{equation}
in agreement with (\ref{sinst}).
 
\subsection{Vertex operators for gauge fields and instanton moduli}
\label{app:vert}

In this subsection we list the vertex operators of the various fields and moduli
of our model, including their normalizations which we express 
in terms of the unit of length $(2\pi\alpha')^{\frac{1}{2}}$. 
In our conventions the vertex operators are always dimensionless and, in general, we
assign canonical dimensions to their polarizations, namely dimensions of (length)$^{-1}$ 
to bosonic fields and dimensions of (length)$^{-{3}/{2}}$ to fermionic ones. However,
in the instanton sector, some of the ADHM moduli acquire different dimensions as indicated
in Tab. \ref{dimensions}. The vertex operators we list in the following are written using the
open string fields appropriate for the D3/D$(-1)$ system, namely the space-time bosonic and fermionic string coordinates $X^\mu$ and $\psi^\mu$, the transverse bosonic and fermionic string coordinates in the complex basis $Z^I$ and $\Psi^I$, the space-time spin-fields $S_\alpha$ and $S^{\dot\alpha}$, 
and the internal ones $S_A$ and $S^A$. Moreover we denote by $\phi$ the bosonic field of
the superghost system. For more details we refer to \cite{Billo:2002hm,Billo:2007py}.

\paragraph{Vertices for gauge fields}
The gauge fields originate from D3/D3 strings. They include a gauge vector $A_\mu$ and a gaugino
$\Lambda^\alpha$ with its conjugate $\bar\Lambda_{\dot\alpha}$ transforming in the adjoint representation, plus bi-fundamental matter fields.
The corresponding vertices at momentum $p$ are
\begin{equation}
\label{gaugevert}
\begin{aligned}
V_A & = (\pi \alpha')^{\frac{1}{2}} \left(A_\mu\right)^{u_A}_{~v_A}\, 
\psi^\mu \ee^{-\phi}\,\ee^{\ii p\cdot X}~,\\
V_{\Lambda}& =  (2 \pi \alpha')^{\frac{3}{4}} 
\left(\Lambda^\alpha\right)^{u_A}_{~v_A}\, S_\alpha S_0 \ee^{-\phi/2}\,\ee^{\ii p\cdot X}~,\\
V_{\bar\Lambda} & = (2 \pi \alpha')^{\frac{3}{4}} \left(\bar\Lambda_{\dot\alpha}\right)^{u_A}_{~v_A}\, S^{\dot\alpha} S^0 \ee^{-\phi/2}\,\ee^{\ii p\cdot X}~,
\end{aligned}
\end{equation}
for the adjoint fields, and
\begin{equation}
 \label{mattervert}
\begin{aligned} 
V_{\phi^I}& =  (\pi \alpha')^{\frac{1}{2}} \left(\phi^I\right)^{u_A}_{~v_{A\otimes I}}\, 
{\bar\Psi}_I \,\ee^{-\phi}\,\ee^{\ii p\cdot X}~,\\
V_{ {\bar\phi}_I}& = (\pi \alpha')^{\frac{1}{2}}  \left({\bar\phi}_I\right)^{u_{A\otimes I}}_{~~\,v_A}\,
\Psi^I \,\ee^{-\phi}\,\ee^{\ii p\cdot X}~,\\
V_{\psi^I}& =  (2 \pi \alpha')^{\frac{3}{4}} \left(\psi^{\alpha I}\right)^{u_A}_{~v_{A\otimes I}}\,
S_\alpha S_I \,\ee^{-\phi/2}\,\ee^{\ii p\cdot X}~,\\
V_{{\bar\psi}_I} & = (2 \pi \alpha')^{\frac{3}{4}} \left({\bar\psi}_{\dot\alpha I}\right)^{u_{A\otimes
 I}}_{~~\,v_A}\, S^{\dot\alpha} S^I \,\ee^{-\phi/2}\,\ee^{\ii p\cdot X}~,\\
V_{F^I}& = {\pi \alpha'} \left(F^I\right)^{u_A}_{~v_{A\otimes I}}\, 
\epsilon_{IJK}\,{\Psi}^J{\Psi}^K \,\ee^{\ii p\cdot X}~,\\
V_{{\bar F}_I}& = {\pi \alpha'} \left({\bar F}_I\right)^{u_A}_{~v_{A\otimes I}}\, 
\epsilon^{IJK}\,{\bar\Psi}_J{\bar\Psi}_K \,\ee^{\ii p\cdot X}~,
\end{aligned}
\end{equation}
for the bi-fundamental matter fields. The internal spin fields appearing in the above formulas are
\begin{equation}
 \begin{aligned}
S^0&=S^{+++}~,~S^1=S^{+--}~,~S^2=S^{-+-}~,~S^3=S^{--+}~,\\
S_0&=S_{---}~,~S_1=S_{-++}~,~S_2=S_{+-+}~,~S_3=S_{++-}~,
 \end{aligned}
\label{spinfields}
\end{equation}
where the $\pm$'s denote the signs of the spinorial weights of $\mathrm{SO}(6)$. Notice that
$S^A$ and $S_A$ trasform in the  representation $R_A$ of the orbifold group.

\paragraph{Vertices for instanton moduli}
Strings with at least one end-point on a D$(-1)$ brane give rise to moduli rather 
than dynamical fields because either they do not have longitudinal Neumann directions at all
or they have mixed boundary conditions. The vertex operators for
D$(-1)$/D$(-1)$ moduli with alike end-points are
\begin{equation}
 \label{D-1vert}
\begin{aligned}
 V_a& = \sqrt{2} g_0 (\pi \alpha')^{\frac{1}{2}}  \left(a_\mu\right)^{i_A}_{~j_A}\,
\psi^\mu\,\ee^{-\phi}~,\\
 V_D& = 2\pi \alpha' \left(D_c\right)^{i_A}_{~j_A}\bar\eta^c_{\mu\nu}\, \psi^\mu\psi^\nu~,\\
 V_M& = \frac{g_0}{\sqrt{2}} (2 \pi \alpha')^{\frac{3}{4}} 
\left(M_\alpha\right)^{i_A}_{~j_A}\, S_\alpha S_0 \ee^{-\phi/2}~,\\
 V_\lambda& = (2 \pi \alpha')^{\frac{3}{4}} \left(\lambda_{\dot\alpha}\right)^{i_A}_{~j_A}\,
S^{\dot\alpha} S^0 \,\ee^{-\phi/2}~,
\end{aligned}
\end{equation} 
where D-instanton gauge coupling $g_0$ is expressed in terms of $\alpha'$ and $g_s$ as in (\ref{g0}).
If the end-points are different we instead have
\begin{equation}
 \label{D-1Ivert}
\begin{aligned}
 V_{\chi^I}& = (\pi \alpha')^{\frac{1}{2}} \left(\chi^I\right)^{i_A}_{~j_{A\otimes I}}\,
 {\bar\Psi}_I \,\ee^{-\phi}~,\\
 V_{ {\bar\chi}_I}& = (\pi \alpha')^{\frac{1}{2}} \left({\bar\chi}_I\right)^{i_{A\otimes I}}_{~\,j_A}\,
\Psi^I \,\ee^{-\phi}~,\\
 V_{M^I}& =  \frac{g_0}{\sqrt{2}} (2 \pi \alpha')^{\frac{3}{4}}
\left(M^{\alpha I}\right)^{i_A}_{~j_{A\otimes I}}\, S_\alpha S_I \,\ee^{-\phi/2}~,\\
 V_{\lambda_I}& = (2 \pi \alpha')^{\frac{3}{4}} \left(\lambda_{\dot\alpha I}\right)^{i_{A\otimes 
I}}_{~\,j_A}\, S^{\dot\alpha} S^I\,\ee^{-\phi/2}~.
\end{aligned}
\end{equation}
The moduli of the charged sector arise from open strings with D3/D$(-1)$ or D$(-1)$/D3 boundary conditions. The vertex operators for those which in the field theory limit describe gauge instantons
are
\begin{equation}
\label{mixgaugevert}
\begin{aligned}
V_w& = \frac{g_0}{\sqrt{2}} (2 \pi \alpha')^{\frac{1}{2}}  \left(w_{\dot\alpha}\right)^{u_A}_{~j_A}\, 
\Delta \,S^{\dot\alpha} \,\ee^{-\phi}~,\\
V_{\bar w}& = \frac{g_0}{\sqrt{2}} (2 \pi \alpha')^{\frac{1}{2}}  \left({\bar w}_{\dot\alpha}\right)^{i_A}_{~u_A}\, \bar\Delta \,S^{\dot\alpha} \,\ee^{-\phi}~,\\
V_\mu& = \frac{g_0}{\sqrt{2}} (2 \pi \alpha')^{\frac{3}{4}} 
\left(\mu\right)^{u_A}_{~j_A}\, \Delta \,S_0 \,\ee^{-\phi/2}~,\\
V_{\bar \mu}& = \frac{g_0}{\sqrt{2}} (2 \pi \alpha')^{\frac{3}{4}} \left(\mu\right)^{i_A}_{~u_A}\, {\bar\Delta} \,S_0 \,\ee^{-\phi/2}~,
\end{aligned}
\end{equation}
while those related to stringy instantons are
\begin{equation}
\label{mixstringvert}
\begin{aligned}
V_{\mu^I} & = \frac{g_0}{\sqrt{2}} (2 \pi \alpha')^{\frac{3}{4}} 
\left(\mu^I\right)^{u_A}_{~i_{A\otimes I}}\, \Delta\, S_I \,\ee^{-\phi/2}~,\\
V_{{\bar\mu}_I}& = \frac{g_0}{\sqrt{2}} (2 \pi \alpha')^{\frac{3}{4}} \left({\bar \mu}_I\right)^{i_{A\otimes I}}_{~u_A}\, \bar\Delta\,S_I \,\ee^{-\phi/2}~,
\end{aligned}
\end{equation}
where $\Delta$ and $\bar{\Delta}$ are the bosonic twist and anti-twist operators respectively and encode the change of boundary condition from Neumann to Dirichlet and \emph{vice-versa}.

\subsection{Derivation of the non-perturbative flux effects for $N_f=N_c-1$}
\label{app:intG}

As discussed in Section \ref{subsecn:Gflux}, the non-perturbative interactions induced by the flux
$G$ are obtained at linear order by computing the integral (see Eq. (\ref{weffg}))
\begin{equation}
{W}_{\mathrm{n.p.}}(G) = \Lambda^{\beta_0}
\int d\,{\widehat{\mathfrak M}}
\,~\ee^{-S^{(0)}_{\mathrm{D3/D(-1)}}(\Phi,\bar\Phi)}\,\left(\frac{\ii}{2}\,G\bar\mu\,\mu\right)
\label{weffg1} 
\end{equation}
where the action on the moduli space is given in Eq. (\ref{sk1}). As we have seen we expect a non-vanishing contribution to this integral when $N_f=N_c-1$, whose schematic form is given in Eq. (\ref{weffG1}).

Here we consider in detail the case of the $\mathrm{SU}(2)$ theory, {\it i.e.} $N_c=2$ and $N_f=1$.
Denoting by $Q^u$ and $\widetilde Q_u$ (with $u=1,2$) the fundamental
and anti-fundamental blocks of the matter superfield $\Phi$ (see Eq. (\ref{Phi})), and by
$\mu'$ and $\bar\mu'$ the non-trivial components of $\mu^1$ and $\bar\mu^1$ respectively,
the moduli action (\ref{sk1}) in this case becomes
\begin{equation}
\begin{aligned}
S^{(0)}_{\mathrm{D3/D(-1)}} =&
~\ii\,D_{c}\big(\bar{w}_{\dot\alpha u}
(\tau^c)^{\dot\alpha}_{~\dot\beta} w^{\dot\beta u}\big)
+\ii \,\lambda_{\dot\alpha} \big(\bar{\mu}_u\,w^{\dot{\alpha} u}
+\bar{w}^{\dot{\alpha}}_{~u}\,\mu^u\big)\\
&~+\left[\frac12\,\bar{w}_{\dot{\alpha}u}\big(Q^u \bar Q_v+\bar{\widetilde Q}^{u}\widetilde Q_v\big)
w^{\dot{\alpha}v}
+\frac{\ii}{2}\,\bar\mu'\,
\bar Q_u\,\mu^u - \frac{\ii}{2}\,\bar\mu_u\,\bar{\widetilde Q}^{u} \,\mu'\right.\\
&\left.\phantom{\frac12}+\ii\,\bar w_{\dot \alpha u}\big(\bar D^{\dot\alpha}\bar{\widetilde Q}^{u}\big)\mu'
-\ii\,\bar\mu'\big(\bar D_{\dot\alpha}\bar Q_u\big) w^{\dot\alpha u}\right]_{\bar\theta=0}
\end{aligned}
\label{smod30}
\end{equation}
where we have explicitly indicated also the two-valued color indices.

The integration over $d^2\lambda$
allows to soak up the two fermionic zero-modes $\mu$ and $\bar \mu$ left after the $G$-flux insertion.
After some elementary algebra, one finds 
a contribution simply proportional to
\begin{equation}
\bar{w}_{\dot{\alpha}u}{w}^{\dot{\alpha}u}~.
\label{contr1}
\end{equation}
To saturate the Grassmannian integrals over $d\mu'$ and $d\bar\mu'$ the only option is to bring down the terms
containing $\bar D_{\dot\alpha}\bar{\widetilde Q}$ and $\bar D_{\dot\alpha}\bar Q$ from the moduli action, thus obtaining a contribution proportional to
\begin{equation}
\left.\big( \bar D^{\dot\alpha}\bar{\widetilde Q}^{u}
\bar D^{\dot\beta}\bar Q_v \big)\right|_{\bar\theta=0}\,
\bar{w}_{\dot{\alpha}u}{w}_{\dot{\beta}}^{~v}~.
\label{contr2}
\end{equation}
Thus, after integrating over all fermionic instanton moduli, we are left with
\begin{equation}
{W}_{\mathrm{n.p.}}(G) = {\mathcal C}\,G\,\Lambda^{5}\,
\big( \bar D^{\dot\alpha}\bar{\widetilde Q}^{u}
\bar D^{\dot\beta}\bar Q_v \big)
\int \!\!d^4w\,d^4\bar w\,d^3D
\,~\ee^{-\ii D_c \bar w \tau^c w -\frac{1}{2}\bar w(Q\bar Q+\bar{\widetilde Q}\widetilde Q)w}
~\bar{w}_{\dot{\alpha}u}{w}_{\dot{\beta}}^{~v}\,
\bar{w}_{\dot{\gamma}r}{w}^{\dot{\gamma}r}
\label{weffg20}
\end{equation}
where we have clumped all numerical constants in the normalization factor $\mathcal C$ and understood that we must set $\bar\theta=0$ in the right hand side.
The bosonic integral (\ref{weffg20}) has been evaluated in Ref. \cite{Matsuo:2008nu} (see
in particular Eq. (5.7) of the published version) and the result is
\begin{equation}
\int \!\!d^4w\,d^4\bar w\,d^3D
\,~\ee^{-\ii D_c \bar w \tau^c w -\bar w A w}
~\bar{w}_{\dot{\alpha}u}{w}_{\dot{\beta}}^{~v}\,
\bar{w}_{\dot{\gamma}r}{w}^{\dot{\gamma}r} = 
\frac{2\epsilon_{\dot\alpha\dot\beta}\,\delta_u^{~v}}{\big(\mathrm{tr}\, A\big)^3}
\label{int20}
\end{equation}
where $A$ is the $2\times 2$ hermitian matrix $A=\frac{1}{2}\big(Q\bar Q+\bar{\widetilde Q}
\widetilde Q\big)$. Exploiting the D-flatness condition (\ref{D-flat1}) for $\xi=0$, it is easy to prove that
\begin{equation}
\mathrm{tr}\, A = \big(\bar M M\big)^{1/2}
\label{AM}
\end{equation}
where $M=\widetilde Q_u Q^u$ is the meson superfield. Using these results in (\ref{weffg20}), after some
simple manipulations and absorbing all numerical factors by redefining the overall coefficient $\mathcal C$, we finally obtain
\begin{equation}
{W}_{\mathrm{n.p.}}(G)= \left. G\,\Lambda^5\,\frac{\bar D^2 \bar M}{(\bar M M)^{3/2}}
\right|_{\bar\theta=0}~.
\label{N2G1}
\end{equation}
As mentioned in the main text, this explicit result is in full agreement with the general expression
(\ref{weffG1}) obtained using dimensional analysis and $\mathrm{U}(1)^3$ charge conservation.

\subsection{Derivation of the non-perturbative flux effects for $N_f=N_c$}
\label{app:intGbar}
Let us now consider the effective interaction induced by a $\bar G$ background flux which, according to Eq. (\ref{weffgbar}), is given by
\begin{equation}
W_{\mathrm{n.p.}}(\bar G) = (\pi\alpha')^2\,\frac{2\pi\ii}{g_s}\,\Lambda^{\beta_0}\,\bar G\, 
\int d\,{\widehat{\mathfrak M}'}
\,~\ee^{-\left.S_{\mathrm{D3/D(-1)}}(\Phi,\bar\Phi)\right.}~,
\label{weffgbar22}
\end{equation}
where in the moduli action it is understood that we have to set $\bar\theta=0$.

We focus on the case $N_0=N_1$, corresponding to SQCD with $N_c=N_f$ flavors. 
The moduli action to be used in (\ref{weffgbar22}) reads explicitly
\begin{eqnarray}
\left.S_{\mathrm{D3/D(-1)}}\right|_{\bar\theta=0} & = & \ii\,D_{c}\big(\bar{w}_{\dot\alpha u}
(\tau^c)^{\dot\alpha}_{~\dot\beta} w^{\dot\beta u}\big)
+\left[\frac{1}{2}\,\bar{w}_{\dot\alpha u}\big(Q^u_{~f} {\bar Q}^f_{~v} + 
{\bar{\widetilde Q}}^u_{~f} {\widetilde Q}^f_{~v}
\big) w^{\dot\alpha v}\right.
\label{actbu} \\
&& \!\!\left.+\frac{\ii}{2}\,\bar\mu^1_f\,
{\bar Q}^f_{~u}\,\mu^u - \frac{\ii}{2}\,{\bar\mu}_u\, {\bar{\widetilde Q}}^u_{~f}\,\mu^{1f}
+\ii\,\bar w_{\dot \alpha u}\big(\bar D^{\dot\alpha}{\bar{\widetilde Q}}\big)^u_{~f}\mu^{1f}
-\ii\,\bar\mu^1_f\big(\bar D_{\dot\alpha}\bar Q\big)^f_{~u} w^{\dot\alpha u}\right]_{\bar\theta=0}
\nonumber
\end{eqnarray}
where $u$ and $f$ are fundamental color and flavor indices respectively.
With respect to Eq. (\ref{sk2}) we
have neglected the quadratic term in $D_c$'s since it appears with an explicit $\alpha'^2$ in front, and it leads to effects of higher order in $\alpha'$ with respect to the ones we will compute in the following.

Since we have chosen $N_f=N_c$, we have the same number of $\mu^u$, ${\bar\mu}_u$ and $\mu^{1f}$, ${\bar\mu}^1_f$
and the integral over these fermionic moduli yields simply
(up to numerical constants which we disregard)
\begin{equation}
 \label{rismm1}
\det {{\bar Q}}\, \det {\bar{\widetilde Q}} = \det \bar M~,
\end{equation}
where $M^f_{~g} = {\widetilde Q}^f_{~u} Q^u_{~g}$ is the meson superfield matrix. 
The above expression has to be evaluated at $\bar\theta=0$, so that only the scalar components appear.

For $N_c=N_f=2$ the integral over the bosonic variables has exactly the form considered in Eq. (5.6) of Ref. \cite{Matsuo:2008nu}:
\begin{equation}
 \label{intw}
 \int d^3D\, d^{4}\bar w\,  d^{4}w\,\ee^{-\ii D_c \bar w \tau^c w -\bar w A w} = \frac{1}{\tr A}~,
\end{equation}
where the $2\times 2$ matrix $A$ is given by
\begin{equation}
 \label{Ais}
A^u_{~v} = \frac{1}{2}\big(Q^u_{~f} {\bar Q}^f_{~v} + {\bar {\widetilde Q}}^u_{~f} {\widetilde Q}^f_{~v}\big)~.
\end{equation}
Using the D-flatness condition (\ref{D-flat1}), it is easy to see that the trace of $A$ can be re-expressed in terms of the low-energy degrees of freedom represented by the meson and baryon superfields $M$, $B$ and $\tilde B$
as follows:
\begin{equation}
 \label{tramb}
\tr A = \big(\tr \bar M M + \bar B B+ \bar {\widetilde B}\widetilde B\big)^{\frac 12}~.
\end{equation}
Inserting into Eq. (\ref{weffgbar22}) the result (\ref{rismm1}) of the fermionic integration and the bosonic integral
(\ref{intw}) we finally get
\begin{equation}
 \label{GbarN2bis}
{W}_{\mathrm{n.p.}} = \mathcal{C} \,\alpha'^2\, \bar G\,\Lambda^{4}\,
\left.\frac{\det \bar M}{\big(\tr \bar M M+\bar B B+ \bar {\widetilde B}\widetilde B\big)^{1/2}} \right|_{\bar\theta=0}
\end{equation}
as reported in (\ref{GbarN2}).

\providecommand{\href}[2]{#2}\begingroup\raggedright\endgroup


\end{document}